\documentclass[prb,groupedaddress,showpacs,twocolumn]{revtex4b5}

\usepackage{graphicx}

\def\figwidth{0.8\columnwidth}

\begin{document}
\title{Competition of random and periodic potentials
in interacting fermionic systems and classical equivalents: the Mott Glass}

\author{T. Giamarchi}
\email{giam@lps.u-psud.fr} \affiliation{Laboratoire de Physique
des Solides, CNRS-UMR 85002, UPS Bat. 510, 91405 Orsay France}
\author{P. Le Doussal}
\email{ledou@lpt.ens.fr} \affiliation{CNRS-Laboratoire de
Physique Theorique de l'Ecole Normale Superieure, 24 rue Lhomond,
75231 Cedex 05, Paris, France.}
\author{E. Orignac}
\email{edmond.orignac@lpt.ens.fr} \affiliation{CNRS-Laboratoire de
Physique Theorique de l'Ecole Normale Superieure, 24 rue Lhomond,
75231 Cedex 05, Paris, France.}
\date{\today}

\begin{abstract}
We study the competition between a random potential and
a commensurate potential on interacting fermionic and bosonic
systems using a variety of methods. We focus on one dimensional
interacting fermionic systems but higher dimensional bosonic and fermionic
extensions, as well as classical equivalents are also discussed. Our
methods which include bosonization, replica variational method,
functional renormalization
group (RG) and perturbation around the atomic limit, go beyond
conventional perturbative
expansions around the Luttinger liquid in one dimension. All these
methods agree on the prediction
in these systems of a phase, the Mott glass,
intermediate between
the Anderson (compressible, with a pseudogap in the optical conductivity)
and the Mott (incompressible with a gap in the optical conductivity)
Insulator.   The Mott glass, which was unexpected from a perturbative
renormalization
group point of view has a  pseudogap in the conductivity while
remaining
incompressible. Having derived the existence of the Mott Glass phase
in one dimension, we show qualitatively that its existence can also be
expected in higher dimension. We discuss the relevance of this phase
to experimental systems such as disordered classical elastic systems
and dirty bosons.
\end{abstract}
%\maketitle must follow title, authors, abstract and \pacs
\maketitle

%\tableofcontents

\section{Introduction}

In many systems, a competition between order and
disorder has drastic consequences for the physical properties.
Such effects are paramount when the pure system has a gap in its
excitations. This situation occurs in a variety of experimental
systems. The most obvious one is a Mott insulator, where
interactions lead to gap in the charge excitations. Low
dimensional systems provide also many experimental situation
where this competition occurs. On the theoretical side, examples
include disordered spin 1 chains
\cite{monthus_spin1_ranexchange}, spin 1/2 ladders with
non-magnetic impurities \cite{azuma_SrCuZnO_af}, disordered Mott
insulators
\cite{kolomeisky_disorder_comm,mori_scba,shankar_spinless_conductivite,fujimoto_mott+disorder_1ch},
doped spin 1 chains \cite{kawakami_dopeds=1}, and disordered
ladder systems
\cite{orignac_2chain_long,orignac_2spinchains,%
orignac_2chain_bosonic,fujimoto_mott+disorder_2ch}. On the
experimental side, examples include doped spin-Peierls systems
\cite{regnault_zinc_doped,grenier_cugesio3}, spin ladder systems
\cite{azuma_SrCuZnO_af}. But such a phenomenon is not limited to
fermionic system. Interacting bosonic systems can also lead to a
Mott insulating phase \cite{haldane_bosons,fisher_boson_loc,giamarchi_attract_1d,giamarchi_mott_shortrev},
with which the disorder can compete. Using
the standard analogy between $d+1$ classical problems and $d$
quantum ones, it is easy to see that such a problem also
encompasses elastic systems such as vortex lines in the presence
of a columnar disorder \cite{nelson_columnar_long,blatter_vortex_review,giamarchi_columnar_variat}.
Other pinned elastic structures such as
the charge density waves \cite{gruner_book_cdw} of spin density waves for which the
competition between a commensurate substrate and disorder easily
occurs are also prime candidates.

In all these systems the disorder tends to close the gap. In some
case the mechanism is simple. Indeed when the ground state is
degenerate and disorder lifts this degeneracy an infinitesimal
disorder causes the formation of domains and leads to a gap
closure \cite{shankar_spinless_conductivite} due to the Imry-Ma
effect. However, in most cases the ground state is not degenerate.
In that case a finite amount of disorder is needed to induce gap
closure\cite{fujimoto_mott+disorder_1ch}. In the latter case, the
complete description of the gap closure is extremely difficult
with the usual analytic techniques such as the renormalization
group due to the absence of a weak coupling fixed point at which
the gap would close. To address this problem one has to tackle
simultaneously strong disorder and strong interactions. A
question of interest is of course what is the nature of such
transition. In particular how one can go from an extremely
ordered (gapped) phase, to the (gapless) disordered one, which is
known to have glassy properties, at least in high enough
dimensions.

Not surprisingly, given the complexity of the problem, very
little is known. In one dimension an RG study combining the
RG for a pure commensurate systems \cite{luther_exact,giamarchi_umklapp_1d}
and an incommensurate
disordered system \cite{giamarchi_loc} has been performed \cite{fujimoto_mott+disorder_1ch}.
Since both the commensurate
potential (Umklapp) and the disorder are relevant operators, no
controlled analysis of the transition could be done. It was
inferred from these studies that one goes from the Mott phase to
the disordered one (Anderson) depending on which operator became
relevant first. The idea of a direct Mott-Anderson transition
seemed the most natural one and was the one usually assumed in
the literature. Solution on a special point (Luther-Emery line)
also supported such conclusions \cite{mori_scba}.

In the present paper we reexamine this problem. For fermionic
systems it is of course difficult to tackle the interactions in
general so we will mostly focus on the one dimensional case
where the interactions can be handled via the bosonization
technique. This allows us to derive a phase Hamiltonian that
makes the connection between this problem and the disordered
elastic problems. A study of this phase Hamiltonian using better suited
methods that capture some non perturbative effects: (i) an
atomic limit; (ii) a variational method; (iii) a functional
renormalization group method, allows to reach the consistent
conclusion that the transition between the Mott insulator and the
Anderson phase (that we call Anderson glass to emphasize its
glassy properties) {\it is not direct}. An intermediate phase, which
has a gap in some of its excitations and yet is glassy does exist. We
determine the characteristics of this phase, that we call the
Mott glass. Since the phase Hamiltonian does describe quantum
crystals and bosons in arbitrary dimension but interacting
fermions only in one dimension, we also give an excitonic
argument directly for the fermions that indicates that this phase
exists also for fermionic systems in dimensions greater than one.
Some of these results were presented in a shorter form in
Ref.~\onlinecite{orignac_mg_short}.

The plan of the paper is as follows. In Sec.~\ref{sec:models} we
introduce fermionic models for the disordered Mott systems. We
then show in Sec.~\ref{sec:bosonization} how in one dimension
this model reduces, using bosonization, to a phase Hamiltonian
that will be the core of our study. Sec.~\ref{sec:elastic} links
this phase Hamiltonian with the other quantum crystals and
classical disordered elastic systems for which our study is
relevant. Sec.~\ref{sec:phasestudy} is devoted to the analysis of
this phase Hamiltonian, using an atomic limit
(Sec.~\ref{sec:phaseatomic}), a variational method
(Sec.~\ref{sec:phasevariational}) and a functional
renormalization group study (Sec.~\ref{sec:phasefrg}). We show
the existence of the Mott glass phase which is both
incompressible and glassy. A reader only interested in the
physical properties of the MG phase can skip these relatively
technical sections and go straight to the
Sec.~\ref{sec:physical_prop} where we examine in details the
physical properties (correlations functions, transport etc.)
Since the link between the phase Hamiltonian and the interacting
fermions only exists in $d=1$ we examine directly the fermions in
higher dimensions in Sec.~\ref{sec:fermionshigh} and give an
atomic limit argument showing that the physics of the Mott glass
phase exists regardless of the dimension. Finally the conclusions
can be found in Sec.~\ref{sec:conclusion}. Some technical
details are pushed in the appendices of the paper.

\section{Models and physical observables} \label{sec:models}

\subsection{Interacting fermions}
We want to study the competition between a Mott and an Anderson
insulator in a dirty fermion system at commensurate filling. The
prototype model for this problem is the extended Hubbard model with a random
on-site potential at half-filling:
\begin{eqnarray}
H &=& -t \sum_{\langle i,j \rangle \sigma} (c^\dagger_{i,\sigma}
c_{j,\sigma} + \text{H. c.}) + U \sum_i n_{i,\uparrow}
n_{i,\downarrow} \nonumber \\
&& + V \sum_{\langle i,j \rangle} n_i n_j + \sum_{i} W_i n_i, \label{eq:dirty_hubbard}
\end{eqnarray}
where $\langle,\rangle$ denotes sum over nearest neighbors, $\sigma$ is the
spin and
and $n_i=n_{i\uparrow}+n_{i\downarrow}$ is the total fermion number on site $i$.
$W_i$ is the random potential at site $i$. For reasons that
will become clear we also include a nearest neighbor repulsion $V$.
A general discussion of the physics of this model will be given in
section \ref{sec:physical_prop}. Given the complexity of this model,
let us first examine a much simpler situation in which explicit
calculations can be performed.

\subsection{Interacting fermions in $d=1$ and bosonized Hamiltonian} \label{sec:bosonization}

In one dimension, many simplifications  occur. Indeed in one dimension,
it is possible to reexpress the Hamiltonian (\ref{eq:dirty_hubbard})
in terms of the  collective charge and spin excitations of the
system. This procedure
is by now standard in one dimension and we refer the reader to
Ref.~\onlinecite{solyom_revue_1d,emery_revue_1d,schulz_houches_revue,voit_bosonization_revue} for more details.
In terms of the bosonic charge $\phi_\rho$ and spin $\phi_\sigma$
collective variables, (\ref{eq:dirty_hubbard}) becomes:
\begin{eqnarray}
H &=& H_\rho+H_\sigma+H_{W} \\
H_\rho &=& \hbar \int \frac{dx}{2\pi} \left[ u_\rho K_\rho (\pi
\Pi_\rho)^2+ \frac{ u_\rho}{ K_\rho  } (\partial_x \phi_\rho)^2\right]
\nonumber \\
&& +\frac{2g_3}{(2\pi a)^2} \int dx \cos \sqrt{8} \phi_\rho \\
H_\sigma &=& \hbar \int \frac{dx}{2\pi} \left[ u_\sigma K_\sigma (\pi
\Pi_\sigma)^2+ \frac{ u_\sigma}{ K_\sigma  } (\partial_x \phi_\sigma)^2
\right] \nonumber \\
&& +\int dx \frac{2g_{1\perp}}{(2\pi a)^2} \cos \sqrt{8} \phi_\sigma \\
H_W &=& \int dx W(x) \rho(x)
\end{eqnarray}
Where $\rho(x)$ is the continuum limit of the charge density and reads:
\begin{eqnarray}\label{eq:density_hub}
&& \rho(x)=-\frac{\sqrt{2} \partial_x \phi_\rho}\pi + \frac{1}{(2\pi\alpha)} [e^{i \sqrt{2}
\phi_\rho -2 k_F x} \cos \sqrt{2} \phi_\sigma \nonumber + \text{h.c.}]\\
&& + \tilde{\rho_0} \cos(\sqrt{8} \phi_\rho -4 k_F x)
\end{eqnarray}
where $\tilde{\rho_0}$ and is renormalized amplitude and $\alpha$ is a
length of the order of the lattice spacing.
All microscopic interactions are absorbed in the Luttinger parameters
$u_\rho,K_\rho$. Spin rotation symmetry leads to $g_{1\perp}=0$ and
$K_\sigma=1$ at low energy.
For very repulsive interactions ($K_\rho < 1/3$) the $4k_F$
density fluctuations are the most relevant, as can be
seen\cite{maurey_wigner} from (\ref{eq:density_hub}). In this
limit\cite{schulz_wigner_1d}, 
 spin fluctuations suppress the $2k_F$ part of the density
fluctuations with repect to the $4k_F$. Let us note that
such limit cannot be achieved within the pure Hubbard model with only
on-site interactions. However, it can be achieved within an extended
Hubbard model with interactions of finite
range\cite{mila_hubbard_etendu,sano_extended_hubbard_1d,schulz_wigner_1d}.  
 Thus one can study a
Hamiltonian containing only the charge degrees of freedom:
\begin{eqnarray} \label{eq:commensurable+desordre_backward}
&& H=\int \frac {dx}{2\pi}\hbar u_\rho \left[K_\rho(\pi\Pi_\rho)^2 +\frac{(\partial_x \phi_\rho)^2}
{K_\rho} \right] \nonumber \\
&& - \frac g {\pi \alpha} \int dx \cos \sqrt{8}\phi_\rho + H_W
\end{eqnarray}
One can perform the rescaling: $\phi=\sqrt{2}\phi_\rho$ and
$\Pi=\Pi_\rho/\sqrt{2}$ which leads to the action,
where we have introduced $v=u_\rho$, $K=K_\rho$
\begin{eqnarray}\label{eq:action_pp+cdb}
&& \frac S \hbar=\int dx \int_0^{\beta \hbar} d\tau \left\{ \frac 1 {2 \pi K}
\left[\frac{(\partial_\tau \phi)^2}v +v
(\partial_x \phi)^2\right] \right.\nonumber \\
&& -\left. \frac{g}{\pi \alpha \hbar}\cos 2 \phi + H_W \right\}
\end{eqnarray}
The Hamiltonian (\ref{eq:commensurable+desordre_backward}) also
describes interacting one-dimensional \emph{spinless} fermions in a
commensurate periodic plus random potential. The lattice form of such
a model is:
\begin{eqnarray}\label{eq:lattice_problem}
H &=& -t\sum_i (c^\dagger_i c_{i+1}+ \text{H. c.} ) +  \sum_i (W_i
+ g \cos(2k_F i a)) c^\dagger_i c_i \nonumber\\
&& +V \sum_i n_i n_{i+1}
\end{eqnarray}
Where $\overline{W_i W_j}=D \delta_{ij}$ and $k_F$ is the Fermi
wavevector of the spinless fermions system. 
In the continuum (\ref{eq:lattice_problem}) leads to:
\begin{eqnarray} \label{eq:fermions-periodic}
H &=& - i  \hbar v_F \int dx \left(\psi_R^\dagger
  \partial_x\psi_R-\psi_L^\dagger \partial_x\psi_L\right) \nonumber \\
&& + V \int dx
  \rho(x)^2 -g \int dx
  (\psi^\dagger_R\psi_L+\psi^\dagger_L\psi_R)\nonumber \\
&& + \int dx W(x) \rho(x)
\end{eqnarray}
$g$ measures the commensurate potential, V the interaction, and
$D$ the disorder strength. Upon bosonization, this Hamiltonian
(\ref{eq:fermions-periodic})  becomes
(\ref{eq:commensurable+desordre_backward}). 
One can thus see that in one dimension,  there
 is no essential difference in the charge sector between a band insulator
(with a $2k_F$
 periodic potential) and a Mott insulator (which can be viewed as a
 system in a $4k_F$ periodic potential
\cite{giamarchi_mott_shortrev}). We stress that in order to establish 
 the equivalence of the bosonized 
charge sector of the Hamiltonian (\ref{eq:dirty_hubbard}) and the
bosonized representation of the Hamiltonian (\ref{eq:lattice_problem})
does not rely on  the equivalence of charge excitations of the Hubbard
model in the limit $U/t \to \infty$ with spinless fermions. 

\subsubsection{Random potential}

Let us now describe qualitatively the effect of the various components
of the random potential.

The effects of disorder on a one dimensional \emph{non-interacting}
system are well known
\cite{berezinskii_conductivity_log,abrikosov_rhyzkin}. Disorder,
however weak localizes all electronic states leading to an insulating
behavior. However, contrarily to what happens in the case of the
periodic potential, there is no gap at the Fermi level but a finite
density of states. Also, the a. c. conductivity does not show a gap but a
behavior of the form $\sigma(\omega)\sim \omega^2 $ up to logarithmic
corrections.
In the presence of interactions, disorder can  be treated by bosonization.
For weak disorder one can separate in the random potential the
Fourier components close to $q\sim 0$ (forward scattering or random
chemical potential) and $q\sim 2k_F$ (backward scattering) as:
\begin{equation}
W(x)=\mu(x) + \xi(x) e^{2i k_F x} + \xi^*(x) e^{-i 2k_F x}
\end{equation}
and treat
them separately. The Hamiltonian
(\ref{eq:commensurable+desordre_backward}) becomes
for $g=0$
\begin{eqnarray}
H_{\text{dis}} &=& \int dx \frac{\hbar v}{2\pi}\left[
K(\pi\Pi)^2+\frac{(\partial_x \phi)^2}{K}
\right] \nonumber\\
&& - \int dx \mu(x) \frac 1 \pi \partial_x\phi(x)
  + \int dx \frac{\xi(x)}{2 \pi \alpha} e^{i 2 \phi(x)} + \text{h.c.} \label{equ:disbos}
\end{eqnarray}
where $\overline{\mu(x)\mu(x)}=D_f \delta(x-x')$ and
$\overline{\xi(x)\xi^*(x')}=D_b \delta(x-x')$ and $D_f=D_b=W$.

The random chemical potential $\mu$  can be  absorbed\cite{giamarchi_loc}  in the
quadratic part of the Hamiltonian (\ref{equ:disbos}) by performing the
transformation:
\begin{equation}\label{eq:simple-transformation}
\phi(x) \to \phi(x)+ \frac{K}{\hbar v}\int^x \mu(x) dx
\end{equation}
Therefore, this term has no role in Anderson localization in the interacting
system, in analogy with the non-interacting case \cite{abrikosov_rhyzkin}.
The backward scattering $\xi$  causes Anderson localization.
Using the renormalization
group \cite{apel_spinless,giamarchi_loc}
it is easy to see that $\xi$ is relevant
for not too attractive  interactions $K < 3/2$ and  becomes of
order one at a length scale
\begin{equation}
l_0 = \alpha \left(\frac{\hbar^2 v^2}{16 W
\alpha K^2}\right)^{1/(3-2K)},
\end{equation}
identified as the localization length in the interacting system.
Beyond this length $l_0$, the phase $\phi$ becomes random and all
correlations decay exponentially.

\subsubsection{Disorder and commensurate potential}

In the absence of disorder a commensurate potential leads to a gap opening
for $K<2$. When disorder is added to such a commensurate phase, its various
Fourier components should be distinguished. Both the forward
scattering and the backward scattering can compete with the
commensurate potential, but as we have seen in the section above, they
can lead to quite different types of ground state. The most
interesting case is the competition of the commensurate potential with
the backward scattering.

In order to understand the competition between the commensurate potential and
the backward scattering, one can argue that the phase physically realized
will be the one with the shortest correlation length. For the Mott phase
the relevant length is the
Mott length $d$, which is the inverse of the gap, or the size of a
charge soliton.  Thus if $d < l_0$ one could expect the system to be a
gapped Mott insulator, whereas for $l_0 < d$ the gap would be washed
out by disorder and the system would be in the Anderson insulating
phase. This qualitative argument can be put on a more formal basis by
writing perturbative RG equations for the coupling constant $g$ of the
commensurate potential and the disorder potential $W$.  Both the pure
commensurate case and the disordered incommensurate one lead to
runaway flow where the coupling constant (resp. $g$ and $W$) reach
strong coupling (resp. at lengths $d$ and $l_0$). A naive
extrapolation consists in assuming that the phase that is physically
realized is the one for which the coupling constant reaches strong
coupling first. Based on such an extrapolation of the RG analysis
\cite{fujimoto_mott+disorder_1ch} one thus expects a
single transition between a commensurate (incompressible) phase and
an Anderson (compressible) insulator.
In order to go beyond this uncontrolled extrapolation to strong coupling
of the RG results, we will use in this paper several non perturbative
methods.

\begin{figure}
\centerline{\includegraphics[angle=0,width=\figwidth]{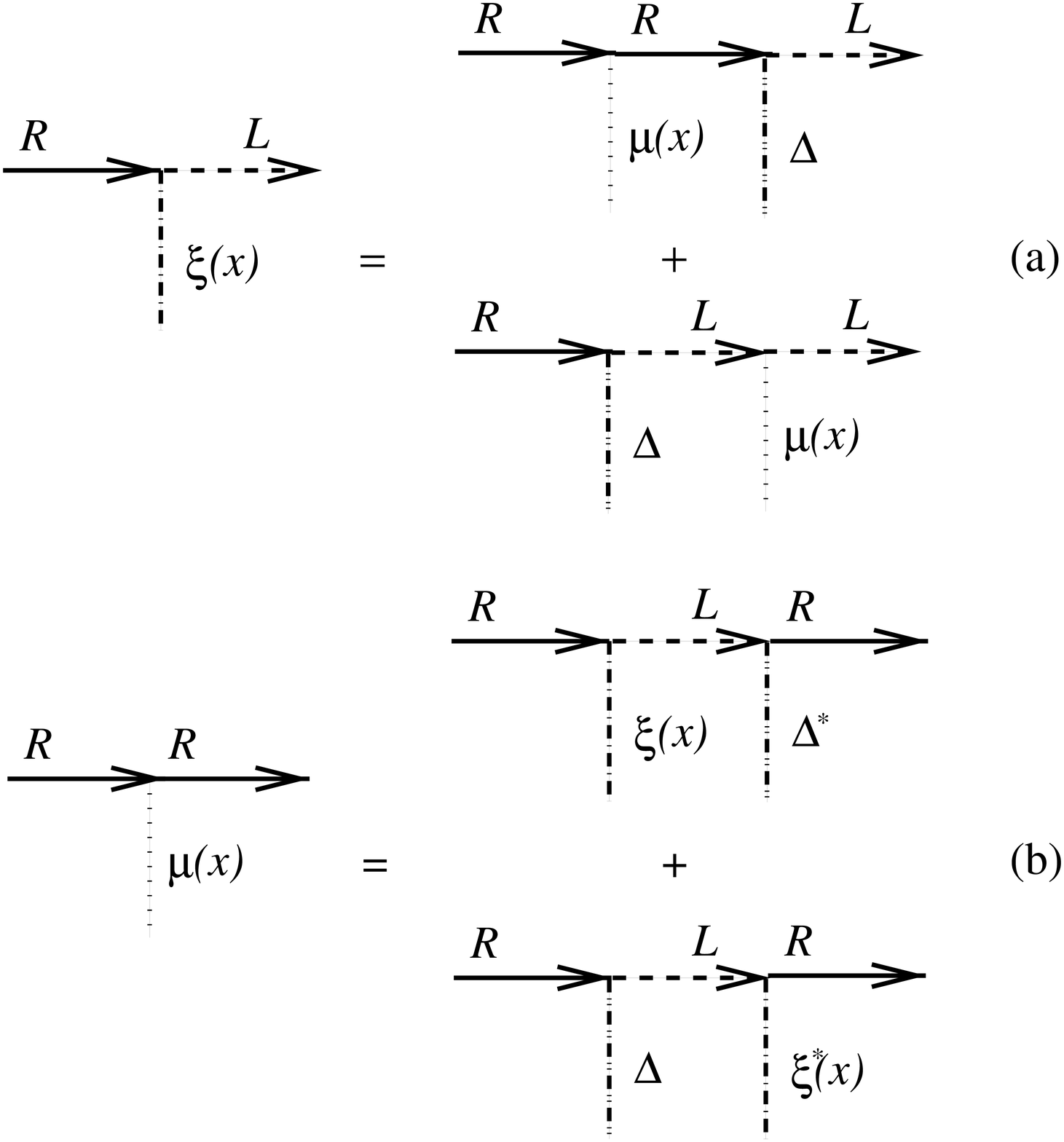}}
\caption{\label{fig:generat}
The generation of effective backward (a) and forward (b) scattering from forward
(resp. backward) scattering and commensurate potential. $\mu$ and $\xi$ denote respectively the forward
and backward part of the random potential (see text) and $g$ the commensurate potential.
$R$ and $L$ denote respectively right and left going fermions with momentum close to $+k_F$ (rep. $-k_F$).}
\end{figure}

Note that a complication arises from the fact that in the presence of
the commensurate potential forward (resp.  backward) random potential
is generated by the backward (resp. forward) component as shown on
Figure~\ref{fig:generat}.
So in principle these two components of the disorder should not be
treated separately.  However, for $K>3/2$ the backward scattering is
irrelevant so one can focus on forward scattering alone. On the other
hand in the limit of strong repulsion, it can be shown that the
closure of the gap that would be induced by a purely forward potential
would occur at much stronger disorder than the one caused by a purely
backward disorder (see Appendix \ref{sec:forward-periodic}). It is
thus reasonable to expect that in the limit of strong repulsion, and
for a backward disorder of the same order of magnitude as the forward
disorder, the closure of the Mott gap is to be attributed to the
backward component of disorder. This allows in the limit of strong
repulsion to neglect the forward component of disorder
altogether. Since in the following we will only consider this case, we
will be justified in dropping the forward component.

A detailed treatment of the forward scattering can be found in
Appendix~\ref{sec:forward-periodic}.

\subsection{Other quantum and classical elastic systems} \label{sec:elastic}

Besides interacting fermions in one dimension, the phase model
describes many other physical systems both in one and in higher
dimension. It is easiest to discuss it in the Lagrangian path integral
formulation. To fix notations let us first generalize the imaginary
time quantum action (\ref{eq:action_pp+cdb}) (entering in the path
integral in $(d+1)$ dimensions as $\int D\phi e^{-S/\hbar}$) to
arbitrary dimension $d$ as follows:
\begin{eqnarray}
&& \frac {\mathcal{S}}{\hbar}= \int_0^{\beta \hbar} d \tau \int d^d x
[\frac{v}{2\pi K}(\partial_x \phi)^2 + \frac 1 {2 \pi v K}
(\partial_\tau \phi)^2 \nonumber \\ && + \frac{V_p(\phi(x))}{\hbar} +
\frac{V(\phi(x,\tau),x)}{\hbar} ] \label{eq:quantumd}
\end{eqnarray}
with $\overline{V(\phi,x)V(\phi',x')}=\delta(x-x') R(\phi-\phi')$. The
bare disorder correlator and bare periodic potential can be chosen as
$R(\phi)=\frac{W}{2 (\pi\alpha)^2} \cos 2\phi$ and $V_p(\phi) = -
\frac{g}{\pi \alpha} \cos(2 \phi)$ respectively, although higher
harmonics, preserving the $\pi$ periodicity, {\it do appear} under
coarse graining and play an important role (even in $d=1$).

Before proceeding, and since the {\it classical limit} of this quantum
action will be of importance below, let us note that it is obtained by
letting $\hbar \to 0$, $K \to 0$, $\overline{K}=K/\hbar$ fixed with
$\beta$ fixed, the zero temperature limit $\beta \to +\infty$ (of
interest here) being taken {\it at the end} (this can be seen e.g. by
rescaling $\tau=\hbar \tau'$ so as to keep the bound of integrations
fixed as $\hbar \to 0$).

There are two types of systems which can be described by
(\ref{eq:quantumd}), quantum elastic systems with point disorder and
classical equivalent systems with {\it correlated disorder} as we now
describe.

\subsubsection{Quantum crystals with point disorder}

Let us consider a quantum crystal in dimension $d$ in a commensurate
periodic potential plus a random potential. In that case, each
particle can be described by its displacement with respect to its
equilibrium position $u(x)$, and the associated phonon modes.  In
general the displacement field has $N$ components, with $N=d$ for
crystals of bosons or fermions, $N=2$, $d=3$ for a crystal of vortex
lines etc.. At $T=0$, the system can still have quantum fluctuations,
leading to a quantum crystal (for a review see Ref.~\onlinecite{giamarchi_quantum_revue}).
Examples of quantum crystals to which
our present study can apply are CDW \cite{gruner_book_cdw},
electron Wigner crystal\cite{chitra_wigner_hall,chitra_wigner_hall_long},
electrons at the surface of helium ($d=2$), stripes in
oxides. Other systems with
$N<d$ such as e.g. a vortex lattice at temperature low enough such
that quantum fluctuations of vortices become important can also be
studied. In that case a periodic potential also exists from the
underlying crystal or, for layered superconductors when the field is
applied parallel to the layers.

The Hamiltonian of such system can be written as: \begin{eqnarray}
H=H_0+H_P+H_W \end{eqnarray} The harmonic part of the Hamiltonian of
the system then reads:
\begin{eqnarray}
H_0=\frac 1 2 \int d^d x \left[ \frac{\Pi_i \Pi_i}{2M} + C_{ij}^{kl}
\partial_{x_i} u_k \partial_{x_j} u_l \right]
\end{eqnarray}
where $x=(x_1,\ldots,x_d)$, the $C_{ij}^{kl}$ is an elastic matrix and
the $\Pi_i$ are the momenta.  The particle density can be written as
the sum\cite{giamarchi_vortex_long}:
\begin{eqnarray}
\rho(x)=\sum_{R} \delta(x- R + u(R)) = \rho_0 \sum_G e^{i G \cdot (x -
u(x))}
\end{eqnarray}
over all reciprocal lattice vectors $G$, and $u=u_1,\ldots,u_d$.  This
allows to write the periodic part of the Hamiltonian as:
\begin{equation}
H=\int d^d x \sum_G V_G e^{i G \cdot u(x)},
\end{equation}
\noindent and the disorder part of the Hamiltonian:
\begin{equation}
H=\int d^d x \sum_G W_G(x) e^{i G \cdot u(x)}
\end{equation}
To minimize technicalities we will not explicitly study the general
case of an arbitrary lattice but a simpler $N=1$ version where one
keeps only one component to the displacement field $u$. In addition of
being already a good approximation in some cases (i.e keeping only the
transverse displacement and its associated shear modulus in a 2d
lattice) the general case is rather similar up to algebraic
complications related to the tensor structure. Thus the quantum
crystal with point disorder and commensurate potential can be modeled
by the quantum action (\ref{eq:quantumd}) with the correspondence
$\phi(x) = \pi u(x)/a$, where $a$ is the lattice spacing, and
$\frac{1}{2 a^2 v \overline{K}}$ the elastic coefficient. We refer the
reader to \cite{chitra_wigner_hall,chitra_wigner_hall_long} for more
detailed descriptions for $N>1$.

\subsubsection{Equivalent classical systems with correlated disorder}

A $d+1$ dimensional classical elastic system in presence of correlated
disorder and periodic potential is described at temperature $T_{cl}$
by its partition sum:
\begin{eqnarray}
Z_{cl} &=& \int D\phi e^{- H_{cl}/T_{cl} } \\ \frac{H_{cl}}{T_{cl}}
&=& \frac 1 {2T_{cl}} [ \int_0^L dz \int d^d x (c(\partial_x \phi)^2 +
c_{44} (\partial_z \phi)^2 \nonumber \\ && + V_p(\phi(x,z)) +
V(\phi(x,z),x)] \label{eq:acclassic}
\end{eqnarray}
where $\phi(x,z)$ is a deformation field and $L$ a thickness in the
direction of correlation.  For a system with internal periodicity,
such as a classical crystal or a classical CDW one has $2 \phi = 2 \pi
u/a$, $a$ being the lattice spacing and $u(x,z)$ the $N=1$
displacement field. In that case the disorder $V(\phi,x)$ and periodic
modulation (i.e the density for a crystal) have the same periodicity
as given above. A prominent example is the flux line lattice in
superconductors (which has $N=2$) in presence of columnar defects. $c$
and $c_{44}$ are then respectively proportional to the bulk (or shear)
and tilt modulus.

The two problems, i.e (\ref{eq:acclassic}) and (\ref{eq:quantumd}) are
thus directly related via the correspondence:
\begin{eqnarray}
z &=& \tau \nonumber \\ L &=& \beta \hbar \nonumber \\ T_{cl} &=&
\hbar \label{eq:corr_class_quant}\\ c_{44} &=& \frac{1}{ \pi v
\overline{K}} \nonumber\\ c &=& \frac {v} { \pi \overline{K}}
\nonumber
\end{eqnarray}
with $\overline{K}=K/\hbar$. The two equivalent models can thus be
studied simultaneously. The classical limit of the $d$ dimensional
quantum model correspond to the zero temperature limit of the $d+1$
equivalent classical model.  Note that the boundary conditions may
differ: periodic for quantum particles (antiperiodic for fermions) but
usually free for the classical system (unless artificially considered
on a torus).  Note also that another correspondence could be defined
with $L=\beta$ and $z=\tau/\hbar$.

\section{Study of the phase model} \label{sec:phasestudy}

In this Section we study the phase model. Before embarking on the
heavy machinery of the replica variational method (in $d=1$) and of
the functional RG (in a $d=4 -\epsilon$ expansion), we first show how
the three phases of the model can be obtained very simply in the
following {\it double limit} (i) classical limit (ii) atomic
limit. Perturbations around those limits can then be done and is not
expected to yield drastic changes, as confirmed by more sophisticated
methods below.

\subsection{phase diagram from the atomic limit} \label{sec:phaseatomic}

We focus in this Section on the {\it classical limit} of the model
(\ref{eq:quantumd}), i.e $\hbar \to 0$, $K \to 0$ with $\overline{K}
\sim K/\hbar$ and $\beta$ fixed and further consider the zero
temperature limit by taking $\beta \to +\infty$ {\it at the end}. We
also perform the rescaling $\phi \to \phi/2$ to simplify the
equations.
 As
will be shown in the following the phases identified here survive at
small enough $K >0$ for $d \geq 1$. Indeed perturbations away from
this limit are irrelevant in the RG sense. As is well known the
classical version is still non trivial since there is still a
competition between the commensurate potential and disorder on one
hand and the elastic term which produces a non trivial classical
configuration satisfying:
\begin{eqnarray}
\frac{\delta S}{\delta \phi^0(x)} &=& - \frac{v}{4 \pi \overline{K}}
\nabla_x^2 \phi^0(x) + \frac{1}{\pi \alpha} ( g \sin(\phi^0(x))
\nonumber \\ && + v(x) \cos(\phi^0(x) - \zeta(x)) ) = 0,
\label{atom}
\end{eqnarray}
where we have define $\xi(x)^*/(\pi a)=iv(x)e^{i\zeta(x)}$.
There may be several solutions to this equation (apart from the global
periodicity $\phi^0(x) \to \phi^0(x) + 2 m \pi$) but the physically
relevant ones that we consider here, are the ones with lowest energy
(or action $S[\phi^0]$) which are selected as $\hbar \to 0^+$.

We now consider the additional limit $\frac{1}{\overline{K}} \to 0$
called atomic limit because the model effectively becomes zero
dimensional in that limit.  In a second stage we describe the
deviations from the atomic limit.  We assume everywhere that the
disorder is {\it bounded} which turns out to be of some importance,
the case of gaussian disorder being discussed later.

\subsubsection{Atomic model: $\overline{K}^{-1}=0$}

Dropping the elastic term in (\ref{atom}) we are thus left to study
the $d=0$ model Hamiltonian in the classical limit
\begin{eqnarray}
H_1(\phi) = - v \cos(\phi - \zeta) - g \cos(\phi)
\end{eqnarray}
with $\zeta$ a random phase distributed uniformly in $[0,2 \pi]$ and
$v$ a random variable, which we choose to have a {\it bounded support}
$-W < v < W$. In the absence of disorder the minima are at $\phi = 2
\pi n$. In the presence of disorder this model has (up to the
periodicity $\phi \to \phi + 2 \pi n$) a unique local (and global)
minimum with probability one for all parameters.  Indeed one can
rewrite:
\begin{eqnarray}
&& H_1(\phi) = - \alpha \cos(\phi - \zeta') \\ && \alpha = \sqrt{v^2 +
2 v g \cos(\zeta) + g^2} \\ && \alpha e^{- i \zeta'} = g + v e^{- i
\zeta}
\end{eqnarray}
and thus there is a single minimum (for $\alpha >0$) at $\phi_0 =
\zeta'$.

An interesting change of behavior however occurs at $W=g$. For $W<g$
the distribution of $\alpha$ is bounded away from $0$ with $g-W <
\alpha < W+ g$ and the new minimum is distributed in the interval $-
\phi_{max} < \zeta'< \phi_{max}$ with $\sin(\phi_{max}) < W/g$. For
$W>g$, $\alpha$ is distributed in the interval $0 < \alpha < W+ g$ and
thus can take values arbitrarily close to zero. Simultaneously, the
new minimum position $\zeta'$ is now distributed in all of $[0,2
\pi]$.

Thus in this simple model two things happen simultaneously as the
disorder width $W$ increases beyond $W=g$. First the distribution of
the Hessian eigenvalue $H''(\phi_0)=\alpha$ extends down to $0$ (while
it is bounded away from zero for weaker disorder) {\it and}, second,
the probability distribution of $\phi_0$ changes abruptly at $W > g$
(while it is bounded in a subinterval of $[-\pi/2,\pi/2]$) below.  As
will become clear below, this abrupt change of behavior corresponds to
a direct transition from the Mott Insulator to the Anderson Glass,
which is in fact a multicritical point in the $\hbar=0$ phase diagram.

\begin{figure}
\centerline{\includegraphics[width=\figwidth]{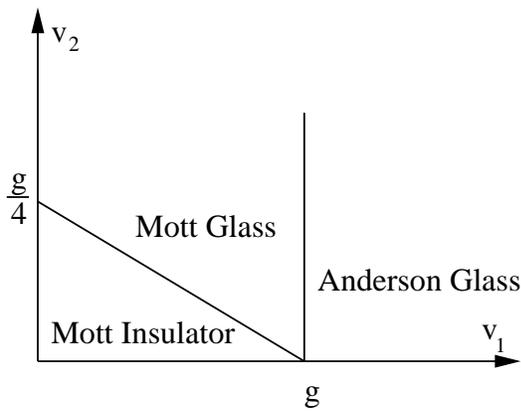}}
\caption{\label{fig:dia1} Phase diagram in the atomic limit: $v_1$ and
$v_2$ are the disorder strengths of the two disorder harmonics and $g$
the strength of the commensurate potential.  The three different
phases are: MI which has a unique local minima with probability $1$;
MG which corresponds to a non zero probability to have two local
minima and zero probability that any minima lie outside of
$]-\pi/2,\pi/2[$; AG which has a non zero probability to have two
local minima and finite probability that minima are outside of
$]-\pi/2,\pi/2[$ (and thus that there are kinks, see text).}
\end{figure}

It turns out that the above form for $H_1(\phi)$ does not yield the
generic behavior for $d=0$. This can easily be seen by adding higher
harmonics and we will illustrate it by simply adding a small second
harmonic to the disorder. It must be stressed that these higher
harmonics are always generated in perturbation theory beyond the
atomic limit (see e.g. Sec.~\ref{sec:phasefrg}) and that they are
generically present in realistic models and should thus be
included. Thus we now study:
\begin{eqnarray}
H_2(\phi) &=& - v \cos(\phi - \zeta) - v_2 \cos(2 \phi - 2 \zeta_2)
\nonumber \\ && - g \cos(\phi)
\end{eqnarray}
with $\zeta_2$ is another random phase uniform in $[0, 2 \pi]$
independent of $\zeta$.  One can rewrite:
\begin{eqnarray}
&& H_2(\phi) = \alpha ( \cos \psi + \beta \cos (2 \psi - 2 \chi) ) \\
&& \psi = \phi - \zeta' \\ && \beta = v_2/\alpha \qquad \chi = \zeta_2
- \zeta'
\end{eqnarray}
and $\alpha$, $\zeta'$ as above. Note that $\chi$ is still uniform in
$[0, 2 \pi]$. Let us consider a fixed $\beta$.  It is easy to see that
for $\beta < 1/4$ there is a unique minimum for any value of $\chi$
and the situation is similar to the one discussed above.  But, as soon
as $\beta > 1/4$ there is a value of $\chi$ for which {\it there is a
second minimum}. At $\beta = 1/4^+$ the second minimum appears for
$\chi=0$.  Thus, new things happen in this model. The phase diagram is
shown in Fig.~\ref{fig:dia1}.
Let us consider for simplicity the case where $v_1 \equiv v$ and $v_2$
are fixed and positive and only the phase are random (the general case
is similar).

\begin{figure}
\centerline{\includegraphics[width=\figwidth]{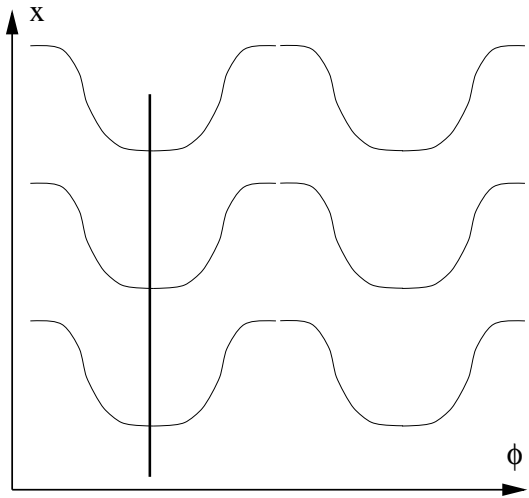}}
\caption{\label{fig:mg1} Pure Mott Insulator phase without disorder. A
ground state in the classical limit $\phi_0(x)=0$ is represented}
\end{figure}

\begin{figure}
\centerline{\includegraphics[width=\figwidth]{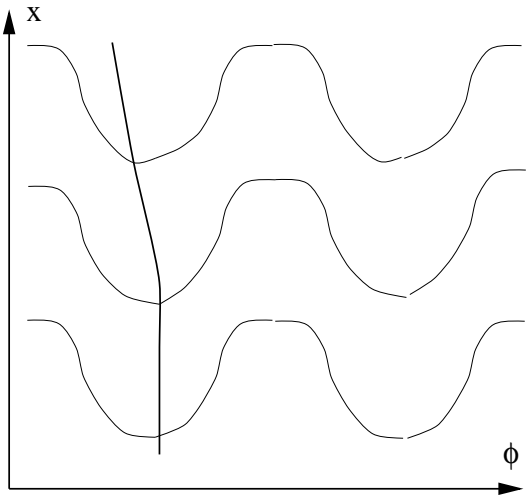}}
\caption{\label{fig:mg2} {\it Mott Insulator phase} with disorder. The
ground state in the classical limit $\phi_0(x)$ is represented. It is
only slightly deformed with respect to $\phi_0(x)=0$. No other local
minima exists (up to the global periodicity $\phi_0(x) \to \phi_0(x) +
2 \pi$). The stability eigenvalues of the Hessian matrix at
$\phi_0(x)$ are strictly positive and bounded from below by a positive
number $m_R^2$. There is a gap $\sim m_R^2$ in the conductivity. The
compressibility is still zero since the response to a tilt $h
\partial_x \phi$ vanishes as no kinks exists}
\end{figure}

There are three phases in this simple, exactly solvable, model:

(i) For $v_1 < g$, $\alpha$ is bounded from below ($\alpha > g - v_1$)
and for small enough $v_2 < \frac{1}{4} (g - v_1)$ a {\it single
minimum} exists: this corresponds to the MI phase as shown in
Fig.~\ref{fig:mg1} and Fig.~\ref{fig:mg2}.

\begin{figure}
\centerline{\includegraphics[width=\figwidth]{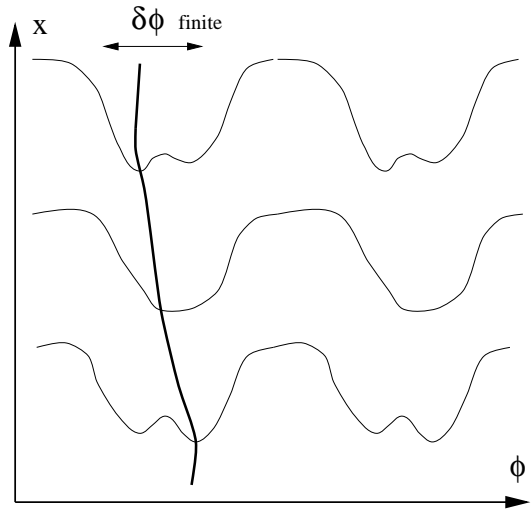}}
\caption{\label{fig:mg3} {\it Mott Glass phase}: the ground state in
the classical limit $\phi_0(x)$ is represented. Other metastable
states exists, and the stability Hessian matrix at $\phi_0(x)$
spectrum extends down to $0$. There is no gap in the conductivity.
The wandering of the ground state along $x$ is bounded as $|\phi_0(x)
- \phi_0(x')|$ is finite (and the $x$ averaged positions is at
$\phi=0$). The compressibility is still zero since the response to a
tilt $h \partial_x \phi$ (i.e a change in chemical potential) vanishes
as no kinks exists between the well separated original minima of the
cosine $\phi=2 n \pi$.}
\end{figure}

(ii) For $v_1 < g$ and $v_2 > \frac{1}{4} (g - v_1)$ {\it two minima
exist}.  This corresponds to the MG phase. Just above the line $v_2 =
\frac{1}{4} (g - v_1)$ the equilibrium position is $\psi = \chi
\approx 0$, $\phi_0 \approx \zeta'$ and $|\phi_0| < \phi_m < \pi/2$
with $\sin \phi_m = v_1/g$. Thus, with probability exactly one the two
minima remain in the wells of the original cosine, i.e the probability
that the new minimum is outside of the interval $[-\pi/2, \pi/2]$ is
exactly zero.  This is represented in Fig.~\ref{fig:mg3}.

\begin{figure}
\centerline{\includegraphics[width=\figwidth]{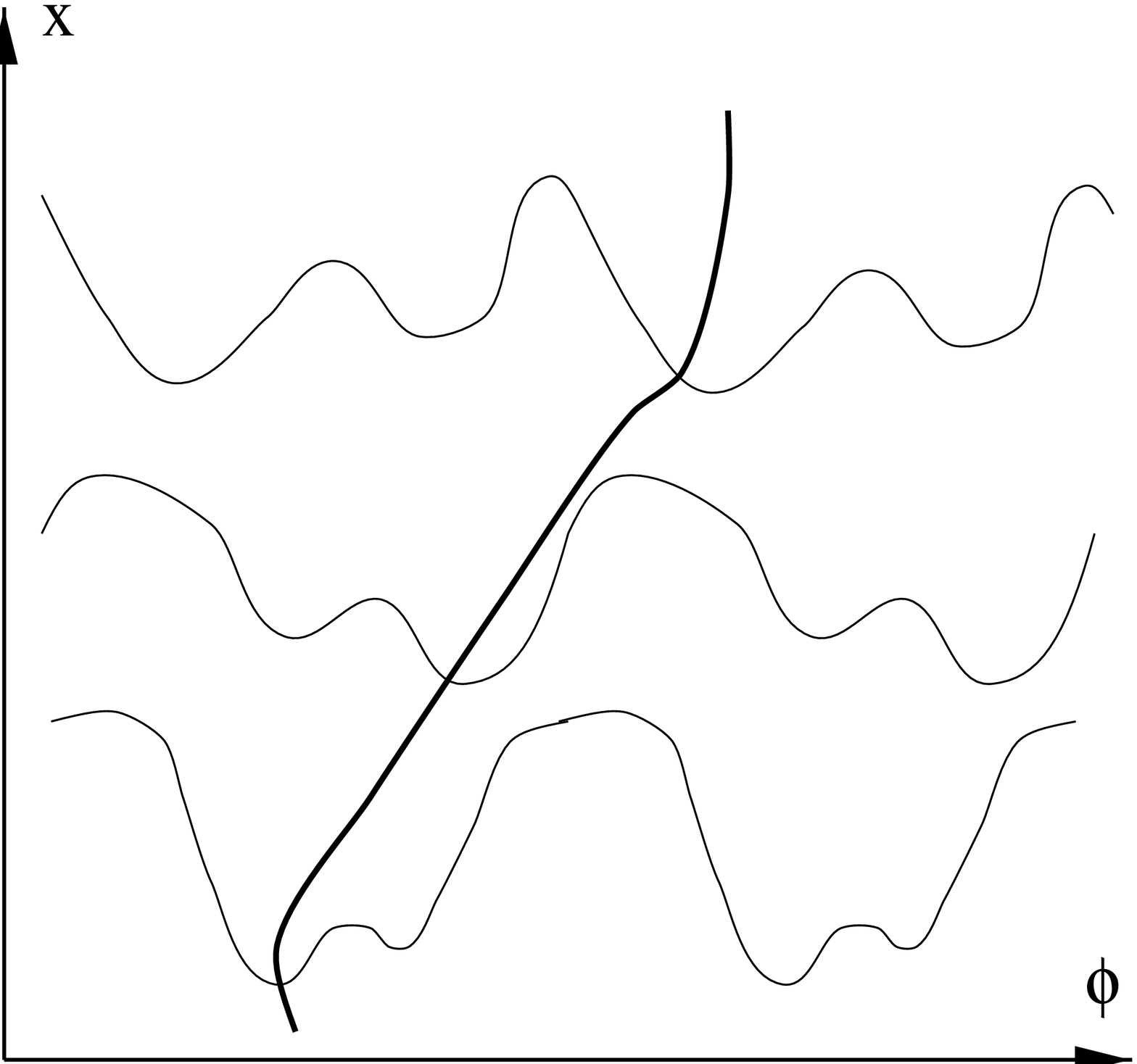}}
\caption{\label{fig:mg4} {\it Anderson Glass phase}: the ground state
in the classical limit $\phi_0(x)$ is represented. Many other
metastable states exist, and the stability Hessian matrix spectrum
extends down to $0$. There is no gap in the conductivity.  The
wandering of the ground state along $x$, $|\phi_0(x) - \phi_0(x')|$ is
unbounded. The compressibility is non zero as the ground state
reorganize in response to a tilt $h \partial_x \phi$ (i.e a change in
chemical potential) because kinks of energies arbitrary close to zero
now exist. }
\end{figure}

(iii) For $v_1 > g$, $\alpha$ has a finite probability to be
arbitrarily close to zero and one easily sees that the probability of
having two minima is nonzero and the probability that the new minimum
is outside of the interval $[-\pi/2, \pi/2]$ is non zero.  This phase
corresponds to the Anderson glass as is shown on Fig.~\ref{fig:mg4}.

\subsubsection{Expansion around the atomic limit, $d \geq 1$}

We can now expand around the atomic limit and consider large but
finite $\overline{K}$. Let us consider $d=1$ for simplicity, but
similar arguments apply to any $d \geq 1$.  We must construct the
classical configuration of the Hamiltonian:
\begin{eqnarray}
H(\phi) &=& \int_x \frac{c}{2} (\nabla_x \phi)^2 - v \cos(\phi(x) -
\zeta(x)) \nonumber \\ && - v_2 \cos(2 \phi(x) - 2 \zeta_2(x)) - g
\cos(\phi(x))
\end{eqnarray}
with $c=v/\pi \overline{K}$. Assume again for simplicity $v$ and $v_2$
constants.  Let us work in the limit of elastic coefficient $c$ very
small (large but finite $\overline{K}$ or equivalently of a long
correlation length for the independent random phases $\zeta(x)$ and
$\zeta_2(x)$). Then we can think of the model as a succession along
$x$ of $d=0$ models (slices) with different realization of the
disorder, and consider e.g. a discretized version:
\begin{eqnarray}
H(\phi) &=& \sum_n \frac{c}{2} (\phi_{n+1} - \phi_{n})^2 - v
\cos(\phi_n - \zeta_n) \nonumber \\ && - v_2 \cos(2 \phi_n - 2
\zeta_{2,n}) - g \cos(\phi)
\end{eqnarray}
where the $\zeta_n$ and $\zeta_{2,n}$ are independent from slice to
slice.

Let us think of a formal perturbation in the elastic coefficient $c$.
For $c=0$ we know the minima for each slice, analyzed above, noted
$\phi_0^n + 2 \pi k_n$.  For $c>0$, one easily sees that to lowest
(naive) order in $c$, to construct the minimal energy configuration
one must first choose which minima of successive slices to connect
together. Small shift in minima positions and the ensuing changes in
minima energies is formally of higher order (they become relevant in
the MG and AG phase and are discussed below but they do not change our
main argument here).  Thus, at each $x$ slice we must choose whether
to connect $\phi_0^n$ with $\phi_0^{n+1}$ (defined as the minimum in
$[-\pi,\pi]$ of the corresponding $d=0$ model) or to $\phi_0^{n+1} \pm
2 \pi$ (which we call a kink). This is equivalent (formally to lowest
order in $c$) to choosing iteratively the one of the three minima
which minimize the distance $\min_{m=0,1,-1}(|\phi_0^n - \phi_0^{n+1}
+ 2 m \pi|)$.  The way these minima are connected define the three
phases.

(i) In the MI regime defined above ($v_1<g$, $v_2<(g-v_1)/4$)
connecting the minima is obvious and leads to a typical ground state
as represented in Fig.~\ref{fig:mg2}, which defines the MI phase.

(ii) In the MG regime ($v_1<g$, $v_2>(g-v_1)/4$) the key observation
is that connecting the minima remains {\it unambiguous} thanks to the
fact that the distribution of the position of these minima remains
confined with probability $1$ within $[- \phi_m , \phi_m]$ with
$\phi_m < \pi/2$. Indeed as long as $\phi_m < \pi/2$ the distance for
a kink $|\phi_0^n - \phi_0^{n+1} \pm 2 \pi| > 2 \pi - 2 \phi_m$ is
always larger than $2 \phi_m$ the distance to connect two minima
maximally separated. Thus the elastic energy always penalizes
kinks. One thus finds that the string $\phi_0^n$ will remain confined,
for small $c$, within the interval $[- \phi'_m , \phi'_m]$ with
$\phi'_m < \pi/2$.  This is represented in Fig.~\ref{fig:mg3} and
corresponds to the Mott Glass phase.

(iii) In the AG regime ($v_1<g$), in each $d=0$ slice $\phi_0^n$ can
be in any position of $[-\pi , \pi]$. Thus connecting the minima
becomes qualitatively (although not quantitatively) the same problem
as for the standard CDW without the periodic potential.  This is the
Anderson Glass phase.

Thus these semi-rigorous arguments show that the three phases which
already exist in the atomic limit {\it do survive} upon adding a small
$c>0$. This also implies that for very small $c$ we expect the phase
boundaries to be close to the ones at $c=0$ (i.e of the $d=0$ atomic
phase diagram Fig.~\ref{fig:dia1}. In practice there are of course
some limitations.  The perturbation in $c$ can be used, strictly, only
in the MI. In the MG perturbation in $c$ cannot strictly apply as the
effect of $c$ on $\phi_0$ is of order $c/\alpha$ and thus generate
terms as $c^2/\alpha$ in the energy, which dominate over $c$ for small
$\alpha$.  Thus, as soon as the Hessian eigenvalues extend to $0$
perturbation theory fails. In the MG this just means that because of
possible multiple minima the ground state cannot be determined
perturbatively. This however only concerns the precise position of
$\phi^0_n$ in the well, but does not have any consequence on the fact
that all $\phi^0_n$ remain in a single well. These effects which go
beyond perturbation theory will change the precise dependence of the
boundaries as a function of $c$ from a naive perturbative estimate,
but we believe that the boundaries are {\it continuous} as $c \to
0$. The $d=0$ phase diagram should thus also give approximately the
$d=1$ one for small $c$.

A useful approximation, used in Sec.~\ref{sec:phasefrg} to study the
MG/AG transition and which consists in replacing the interaction term
by a mass $\frac{1}{2} m^2 \phi^2$, can be checked in the atomic
limit. It gives correctly the multicritical point at $v=m^2$ since it
has a transition for $W=m^2$ (for $W< m^2$ there is a unique minimum
with probability $1$ while for $W> m^2$ there is a finite probability
that there is a second minimum).  By construction this approximate
model cannot distinguish between the MG and the AG phase, and is only
used to describe the MI/MG transition.

One must emphasize that none of the transition in the $d=0$ model in
Fig.~\ref{fig:dia1} survives if the local distribution of disorder is
gaussian. Indeed the probability of a second minimum is always non
zero, although it can be exponentially small, resulting in sharp
crossover behavior rather than transitions. It is however not so clear
whether the same apply in higher dimension. We know that the gap is
robust to small bounded disorder. However for unbounded disorder,
formation of terraces if size $L$ become energetically favorable in
rare regions with exponentially small probability. This leads to a
gapless spectrum, with an exponentially (in $d>1$) small density of
state at low energy.  Note that although the gap itself cannot be used
anymore as an order parameter one can still clearly define a phase
transition between MI and MG since in the MG phase the spectrum is
expected to become algebraic.

To conclude this Section we have established that for bounded disorder
three phases (MI, MG, AG) already exist in the atomic classical limit.
Previous attempts at analyzing the classical limit
\cite{fukuyama_disorder+commensurate} assumed that beyond a length
$L_0$ the distribution of the phase becomes random.  As we find here,
this is incorrect for weak disorder and the phase has instead a narrow
distribution around $\phi=0$. Thus these arguments missed the
existence of the Mott Glass phase and the correlation length $L_0$
identified in Ref. \onlinecite{fukuyama_disorder+commensurate} is not
the correct one. In order to get more detailed information on the
three phases in $d=1$ and higher we now turn to more sophisticated
methods.

\subsection{Replica Variational Method}\label{sec:phasevariational}

Let us now use a Gaussian variational
method\cite{mezard_variational_replica,giamarchi_columnar_variat} to study the action
 originating from the Hamiltonian
(\ref{equ:disbos}). We only retain in this section the backward
scattering which is the one leading to localization. The effect of
forward scattering is analyzed in appendix~\ref{sec:forward-periodic}.
Let us recall that backward scattering is generated even if only the
commensurate and forward scattering are present.

\subsubsection{Derivation of the variational equations}\label{sec:variational_backward}

Let us first recall briefly the principle of method we use.  The
partition function of a disordered quantum system is
\begin{equation} \label{eq:partition_function}
Z=\int d[\phi] e^{-\frac{S[\phi]}{\hbar}}
\end{equation}
where $S[\phi]$ is the Euclidean action that depends explicitly on the
quenched disorder. In physical problems, one usually needs the average
of free energy
\begin{equation}
\label{eq:average_free_energy}
F=-\frac 1 \beta \overline{\ln Z}
\end{equation}
where the overbar denotes a disorder average.  This average can be
done via the replica trick\cite{edwards_replica}:
\begin{equation} \label{eq:replica_trick}
\ln Z=\lim_{n \to 0} \frac{Z^n-1}{n}
\end{equation}
One has for integer n:
\begin{equation} \label{eq:z_puissance_n}
\overline{Z^n}=\int \overline{\prod_{a=1}^n d[\phi_a] e^{-\frac
{S_a[\phi_a]}{\hbar}}}=\int \prod_{a=1}^n d[\phi_a]
e^{-\frac{S_{\text{rep.}}[{\phi_a}]}{\hbar}}
\end{equation}
where $S_{\text{rep.}}$ is a disorder free quantity that depends on
the $n$ fields $\phi_a$. In the end, one has traded the disorder
average to a analytical continuation from an integer number of fields
$n$ to $n=0$.  In our particular case, after having averaged over
disorder the action (\ref{eq:action_pp+cdb}) leads to the replicated
action:
\begin{widetext}
\begin{eqnarray}\label{eq:pp+cdb_replica}
\frac{S_{\text{rep.}}}{\hbar}&=&\sum_a \left[ \int \frac{dx
d\tau}{2\pi K} \left(v (\partial_x \phi_a)^2 + \frac{(\partial_\tau
\phi_a)^2}{v}\right) -\frac {g}{\pi\alpha \hbar} \int dx d\tau \cos 2
\phi_a \right] \nonumber \\ &-& \frac {W}{(2\pi\alpha \hbar)^2}
\sum_{a,b} \int dx \int_0^\beta d\tau \int_0^\beta d\tau' \cos
\left(2(\phi_a(x,\tau)-\phi_b(x,\tau'))\right)
\end{eqnarray}
\end{widetext}
in which one has to take the limit $n \to 0$.  One way to perform this
limit is to use a Gaussian Variational Method (GVM) initially
introduced to study classical disordered systems such as random
heteropolymers \cite{shakhnovich_variational}, random manifolds
\cite{mezard_variational_replica}, and vortex lattices
\cite{giamarchi_vortex_long,giamarchi_columnar_variat}.  This Ansatz
has been extended to treat correlated disorder and thus to apply to
quantum systems as well \cite{giamarchi_columnar_variat}.  Since this
method has been shown to describe with a good accuracy both the pure
commensurate phase and the Anderson insulator, on can expect to get
also good results in this more complicated situation.

The ansatz consists in finding the ``best'' quadratic action
\begin{equation}\label{eq:variational_action}
S_0 = \frac 1 {2 \beta \hbar} \sum_{\omega_n} \int \frac{dq }{2\pi}
\phi_a(q,\omega_n) G_{ab}^{-1}(q,\omega_n)\phi_b(q,\omega_n)
\end{equation}
with :
\begin{equation}
v G^{-1}_{ab}(q,\omega)=\frac {((vq)^2+\omega^2)}{\pi K}\delta_{ab} -
\sigma_{ab}(q,\omega)
\end{equation}
i.e. the one that minimizes the trial free energy:
\begin{equation} \label{equ:varbase}
F_{\text{var}}=F_0 +\frac 1 {\beta \hbar} \langle
S_{\text{rep.}}-S_0\rangle_{S_0}
\end{equation}
Where $\langle \ldots \rangle_{S_0}$ designates the averages performed
with respect to the Gaussian action and $F_0$ is the free energy
associated with (\ref{eq:variational_action}) i.e.
\begin{equation}
  \label{eq:free_enegy_gaussian} F_0= \int \frac{dq}{2\pi} \frac 1
\beta \sum_n (\ln G)_{aa}(q,\omega_n)
\end{equation}
In this method the full Green's functions $G(q,\omega)$ are the
variational parameters. The derivation of the saddle point equations
is performed in appendix~\ref{sec:appendix_gvm}. Introducing
$G_c(q,\omega_n) = \sum_b G_{ab}(q,\omega_n)$,
$B_{ab}(x,\tau)=G_{aa}(x,\tau)-G_{ab}(x,\tau)$ and parameterizing these
$n\times n$ hierarchical matrices using Parisi's ansatz ($G_{a\ne b}
\to G(u)$, $G_{aa}\to\tilde{G}$, and similarly for any quantity, see
appendix~\ref{sec:appendix_gvm}), we obtain the saddle point equations
(\ref{saddle-point-backward-appendix}):
\begin{widetext}
\begin{eqnarray}\label{saddle-point-backward}
G_c^{-1}(q,\omega_n) & = & \frac \hbar {\pi K} (v
q^2+\frac{\omega_n^2} v) +\frac{4 g }{\pi \alpha } \exp\left(-2 \hbar
\tilde{G}(x=0,\tau=0)\right) \nonumber \\ &+ &\frac{2W}{\hbar(\pi
\alpha )^2}\int_0^{\beta \hbar}d\tau (1-\cos(\omega_n \tau))
\left[\exp(-4\hbar \tilde{B}(x=0,\tau))-\int_0^1 du \exp (-4 \hbar
B(u))\right]
\end{eqnarray}
\end{widetext}
with
\begin{eqnarray}
\sigma(q,\omega_n,u)=\frac{2 W v }{ (\pi \alpha )^2}\beta \exp (-\hbar
4 B(u))\delta_{\omega_n,0}
\end{eqnarray}
In the absence of the commensurate potential,
(\ref{saddle-point-backward}) are known to lead to a one step replica
symmetry broken solution describing the Anderson localized
phase\cite{giamarchi_columnar_variat}, the replica symmetric solution
being unstable.  On the other hand, in the absence of disorder the
commensurate phase (i. e. the Mott Insulator) is obviously replica
symmetric. Therefore, we will search first for a replica symmetric
solution that we expect to be associated with a Mott insulating phase
in Sec.~\ref{sec:backward_rs}. Above a certain disorder this solution
will become unstable, and one has to turn to replica symmetry broken
solutions. As we will see in Sec.~\ref{sec:rsb_solution}, in the
presence of the commensurate term, there is, besides the saddle point
solution corresponding to the Anderson insulator, there is room for a
third saddle point solution corresponding to the Mott glass phase.

\subsubsection{Gapped replica symmetric solution: Mott Insulator}\label{sec:backward_rs}

For the replica symmetric solution, $G(q,\omega_n,u)=G(q,\omega_n)$.
The saddle point equations then read:
\begin{widetext}
\begin{eqnarray}
v G_c^{-1}(q,\omega_n) &=& \frac \hbar {\pi K} ( (v q)^2+\omega_n^2) +
m^2 +I(\omega_n) \label{eq:saddle_point_RS_Gc} \\
G(q,\omega_n,u) &=& \frac{2 W\beta
K^2}{(\hbar \alpha )^2} \frac{e^{-4 \hbar
G_c(0)}\delta_{\omega_n,0}}{((vq)^2+ \pi \overline{K} m^2)^2}
\label{eq:saddle_point_replica_symmetric} \\
m^2 &=& \frac{4 g v }{\pi \alpha}e^{-2\hbar\tilde{G}(0,0)} \\
I(\omega_n) &=& \frac{ 2 W v}{(\pi\alpha)^2 \hbar }e^{-4 \hbar
G_c(0,0)}\int_0^{\beta \hbar}d\tau (e^{4 \hbar G_c(0,\tau)}-1)(1-\cos
(\omega_n \tau)) \label{sigma_omega_RS}
\end{eqnarray}
\end{widetext}
In the general case, the equations
(\ref{eq:saddle_point_RS_Gc})--(\ref{sigma_omega_RS}) can
only be solved numerically. However, in the limit $\hbar \to 0$ , $K
\to 0$, $\overline{K}=K/\hbar$ fixed, it is possible to solve these
equations analytically. This is due to the fact that in that limit
$m^2$ has no dependence on $I(\omega_n)$, so that $m$ is given by the
simple equation:
\begin{equation} \label{sigma_omega=0_RS}
m^2=\frac {4 g v}{\pi \alpha} \exp \left[ -\frac{ W \overline{K}^{1/2}
}{ \alpha^2 \pi^{3/2} v^{1/2} m^3} \right]
\end{equation}

The self-consistent equation (\ref{sigma_omega=0_RS}) has always the
trivial solution $m=0$. Let us determine under which conditions it can
also have a non-trivial solution with $m \neq 0$. In order to obtain
the answer in physical terms it is convenient to reexpress all
quantities as a function of the physical lengths  $l_0$ and $d$:
\begin{eqnarray}
\frac1{l_0^3}=\frac{16 W \overline{K}^2}{(\alpha v)^2},
\label{eq:definition_l0} \\
\frac1{d^2}=
\frac{4 g \overline{K}}{(\alpha v)}\label{eq:definition_d}.
\end{eqnarray}
They correspond respectively to the localization (pinning) length in the absence of
commensurability, and to the soliton size of the pure gap
phase.  We introduce the length
\begin{equation}
\xi^2 = \frac{v^2}{(\pi \overline{K} m^2)}
\end{equation}
which, as we will see, is the correlation length in the presence of both
the commensurate potential and the disorder.  One can then rewrite
(\ref{sigma_omega=0_RS}) as:
\begin{equation}\label{eq:m_reduced}
\frac 1 {\xi^2}=\frac 1 { d^2}\exp\left[-\frac 1 {16} \left( \frac \xi
{l_0}\right)^3\right]
\end{equation}
For $l_0/d >\frac 1 2 \left( \frac {3e} 4 \right)^{1/3}$, this
equation admits two solutions. It can be seen by considering the limit
of large disorder that solutions with $l_0/\xi < \frac 1 2 \left(\frac
3 4\right)^{1/3}$ are spurious. Thus, for $l_0/d >\frac 1 2 \left(
\frac {3e} 4 \right)^{1/3}$, we have a unique solution of
(\ref{eq:m_reduced}).  For $l_0/d <\frac 1 2 \left( \frac {3e} 4
\right)^{1/3}$, there is no solution, which means that the replica
symmetric solution with a mass is unstable. Since it is known
\cite{giamarchi_columnar_variat} that a replica symmetric phase with
no mass is unphysical for $\hbar \to 0$, this means that for large
enough disorder, we obtain a breaking of replica symmetry.

Let us now examine the equation for $I(\omega_n)$.  An expansion
around $\hbar=0$ in equation (\ref{sigma_omega_RS}) gives the
self-consistent equation for $I(\omega_n)$:
\begin{equation}
I(\omega_n)=\frac{ 8 W v}{ (\pi \alpha)^2}\int_0^{\beta \hbar}
G_c(x=0,\tau)(1-\cos(\omega_n \tau)) d\tau
\end{equation}
Going to Fourier space and performing the $q$ integration leads to to
the final form:
\begin{equation}
I(\omega_n)= \frac{4 W v \overline{K} }{ \pi \alpha^2}\left[
\frac{1}{\sqrt{\pi \overline{K}}m} -\frac 1 {\sqrt{\omega_n^2 +\pi
\overline{K} (m^2+I(\omega_n))}} \right]
\end{equation}
Obviously,
\begin{equation}
I(\omega_n)= m^2 f\left(\frac{\omega_n}{\sqrt{\pi
\overline{K}}m}\right)
\end{equation}
where $f$ satisfies the equation:
\begin{equation} \label{equation-for-f}
f(x) = \lambda \left[ 1-\frac 1 {\sqrt{1+x^2+f(x)}}\right]
\end{equation}
and
\begin{equation}\label{eq:def-lambda}
\lambda=\frac{4 W \overline{K}^{1/2} v }{ \pi^{3/2} \alpha^2
m^3}=\frac 1 {4} \left(\frac \xi {l_0}\right)^{3}
\end{equation}
As can be seen from (\ref{eq:saddle_point_RS_Gc}), $m$
defines the correlation length $\xi$ in the presence of \emph{both}
the commensurate and the random potential.

Once $m \neq 0$ is known from (\ref{sigma_omega=0_RS}), $\lambda$ and
therefore $f$ and $I(\omega_n)$ are entirely determined. The above
equations thus completely fix all the parameters of the gapped RS
phase.

\begin{figure}
\centerline{\includegraphics[width=\figwidth]{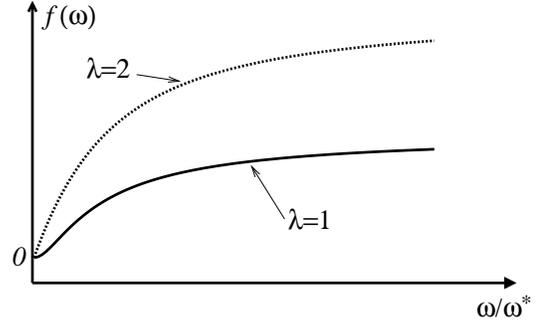}}
\caption{\label{selfenergy} The function $f$ as a function of $\omega$
for $\lambda=1$ (full curve) and $\lambda=2$ (dashed curve). Note that
as $\lambda=2$, $f$ starts linearly with frequency, contrarily to
$\lambda < 2$ for which $f$ is quadratic at small $\omega$.}
\end{figure}

For $\lambda <2$, there is a physical solution of
(\ref{equation-for-f}) such that $\lim_{x \to \pm \infty} f(x)=
1+\lambda$ and for $x \ll 1$ , $f(x)=1+\alpha x^2 +o(x^2)$ with
$\alpha=\lambda/(2-\lambda)$. The corresponding behavior of $f(x)$ as
a function of $\omega$ is shown on Fig.~\ref{selfenergy} for
$\lambda=1$.
For $\lambda=2$, for $x \ll 1$, $f(x)=1+\frac 2 {\sqrt{3}} \mid x \mid
+O(x^2)$ and $\lim_{x \to \infty} f(x)=3$.  The corresponding graph of
$f(x)$ is also shown on Figure~\ref{selfenergy}.
 For $\lambda>2$
(\ref{equation-for-f}) has no physical solution.
Thus, $\lambda=2$ is  the
boundary for the gapped RS phase.  Putting $\lambda=2$
in (\ref{eq:def-lambda}) leads to $l_0/\xi=\frac 1 2 >\frac 1 2
\left(\frac 3 4\right)^{1/3}$.  Reinjecting this value in
(\ref{eq:m_reduced}) gives
\begin{equation}\label{eq:condition_marginal}
 \frac {l_0}{d}= \frac 1 2 e^{\frac 1 {4}}
\end{equation}
This point is in the domain where (\ref{eq:m_reduced}) still has
solutions.
Therefore, as disorder increases the system
attains the point where $\lambda=2$ \emph{before} reaching the point
where $m=0$.  Beyond the point $\lambda=2$ the replica symmetric
solution becomes unstable. This leads us to consider a replica
symmetry breaking solution of Eq.~(\ref{saddle-point-backward}) for
$\frac {l_0}{d} < e^{\frac 1 {4}}/2$  allowing  for a non-zero
$m$. The corresponding phase will thus not be the simple Anderson
localized phase expected from the strong coupling RG argument\footnote{In a previous publication,
we have used a slightly different definition for $l_0$ which lead to different
values of $l_0/d$.}.

\subsubsection{Replica symmetry broken solution}\label{sec:rsb_solution}

In the preceding section, we have seen that in the limit $\hbar \to 0$
a replica symmetric solution of equations
(\ref{saddle-point-backward}) can exist for $\lambda \leq 2$ but is
unstable for $\lambda>2$. In this section, we consider in the limit
$\hbar \to 0$ the one-step replica symmetry breaking (RSB) solution of
equations (\ref{saddle-point-backward}). Such RSB solution should be
valid for $\lambda>2$, and is known to correctly describe the simple
Anderson insulating phase for the simple disordered
case\cite{giamarchi_columnar_variat}.  Compared to the case in the
absence of commensurate potential we have to allow, in our RSB
solution, for $m^2 \neq 0$. However, contrarily to the RS case, a RSB
solution with $m^2 = 0$ is perfectly possible
\cite{giamarchi_columnar_variat} and corresponds to the case in which
the commensurate potential is completely washed out by the random
potential.

Two scenarios could thus be a priori possible. Either as soon as the
replica symmetric solution becomes unstable one obtains a RSB solution
with $m^2=0$ similar to the solution of
Ref.~\onlinecite{giamarchi_columnar_variat} or there can exist an
intermediate regime with both RSB and $m^2 \neq 0$.  The first case
would correspond to the simple scenario, suggested from extrapolating
the RG of a direct transition between the commensurate phase and the
Anderson insulator. On the other hand the behavior in the RS solution
strongly suggests the existence of an intermediate phase, the Mott
Glass. Indeed in the RS phase the optical gap closes, leading to a
conductivity much similar to the simple Anderson one, whereas the
compressibility remains zero. This suggests that all effects of the
commensurate potential have not yet disappeared, and that the system
is not in a simple Anderson regime.  As we will see, this is this
second possibility that obtains, leading thus to a much richer
physical behavior than could have been guessed from the RG
extrapolations.

The saddle point equations (\ref{saddle-point-backward}) are first
rewritten as:
\begin{widetext}
\begin{eqnarray} \label{eq:gcrsb}
v G_c^{-1}(q,\omega_n)&=&\frac 1 {\pi \overline{K}}((v
q)^2+\omega_n^2) + m^2+\Sigma_1(1-\delta_{n,0}) +I(\omega_n) \\
I(\omega_n)&=&\frac{2 W v}{ (\pi \alpha )^2 \hbar }\int_0^{\beta\hbar}
\left[e^{-4\hbar \tilde{B}(\tau)}-e^{-4\hbar
B(u>u_c)}\right](1-\cos(\omega_n \tau)) d\tau \\
\Sigma_1&=&u_c(\sigma(u>u_c)-\sigma(u<u_c))=[\sigma](u>u_c) \\
\sigma(u)&=&\frac{2 W v}{ (\pi \alpha)^2}e^{-\hbar 4 B(u)}\beta
\delta_{n,0} \\ m^2 &=& \frac{4 g v}{\pi \alpha}e^{-4 \hbar \tilde G
(0)}
\end{eqnarray}
\end{widetext}
where we look for a one step RSB solution, as is adapted to
$d=1+1$. The parameters $m$, $\Sigma_1$ and the breakpoint $u_c$ have
to be determined self-consistently. The full solution is given in
Appendix~\ref{sec:appendix_rsb}. The parameter $u_c$ is determined
from the marginality of the replicon condition, which has been shown
\cite{giamarchi_columnar_variat} to be the correct condition to
impose. This leads to $I(\omega_n) \propto \mid \omega_n \mid$. One
can check (see Appendix~\ref{sec:appendix_rsb} and
Sec.~\ref{sec:backward_rs}) that this condition is also satisfied by
$I(\omega_n)$ at the limit of stability of the RS solution. The two
other parameters $m$ and $\Sigma_1$ depend on the ratio $d/l_0$.

As is shown in Appendix~\ref{sec:appendix_rsb} for $d/l_0 > 1.86$
one has $m=0$ and $\Sigma_1 \ne 0$. This solution is thus similar to
the one of a disordered system without the commensurate potential and
corresponds to an Anderson glass (insulator).  Such a phase has no gap
in the optical conductivity \cite{giamarchi_columnar_variat}.  Since
there is no mass in the propagator (\ref{eq:gcrsb}) for $\omega_n=0$
the compressibility is finite.

On the other hand for $d/l_0 < 1.86$ the solution has a finite mass.
This regime is thus different from the simple Anderson insulator. The
physical properties (conductivity, phason density of states,
compressibility) will be discussed at length in
Sec.~\ref{sec:physical_prop}.  An important result of
Sec.~\ref{sec:physical_prop} is that because of the presence of the
mass the system is still incompressible while having the conductivity
of an Anderson Insulator. This is the Mott glass phase
\cite{orignac_mg_short} which shares some properties of the Mott
Insulator (incompressibility) with those of a glassy phase (breaking of
replica symmetry).

\begin{figure}
\centerline{\includegraphics[width=\figwidth]{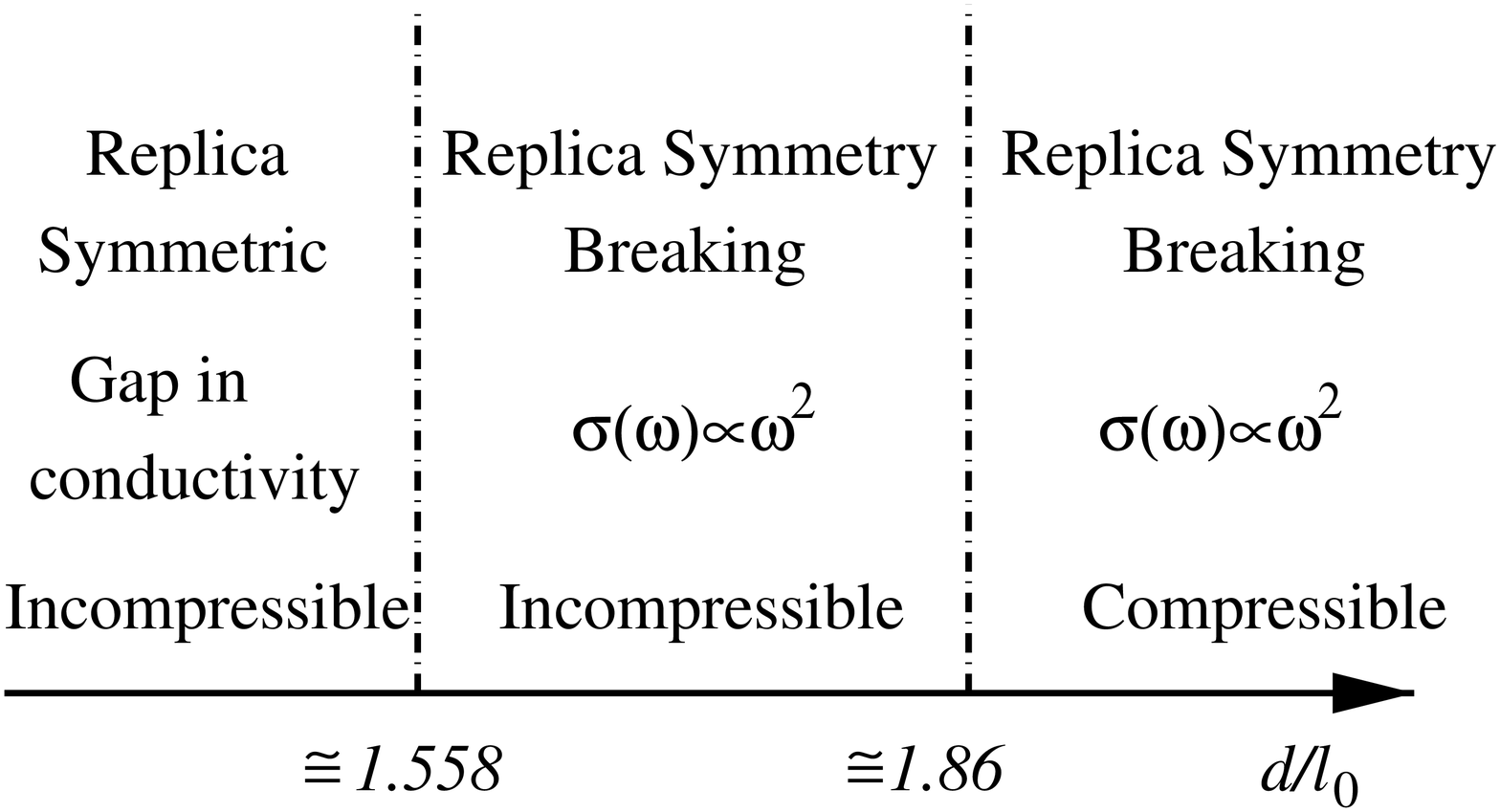}}
\caption{\label{fig:phase_diagram} The phase diagram of the
commensurate system with backward scattering only as a function of
$d/l_0$ (disorder increases when $d/l_0$ increases).
At weak disorder,$d/l_0\ll 1$ one obtains an incompressible
phase with a gap in the conductivity i.e. a Mott Insulator. For strong
disorder, $d/l_0 \gg 1$, one has a compressible phase with a
conductivity that behaves as $\sigma(\omega)\propto \omega^2$ i.e. an
Anderson glass. The surprising feature of this phase diagram is the
appearance of an intermediate incompressible phase (like the Mott
insulator) having the same conductivity as an Anderson glass for
$d/l_0 \sim 1$.}
\end{figure}

Therefore, the physical picture is the following: for $d/l_0<
2e^{-1/4}$, one has a replica symmetric solution with a gap in the
conductivity, the Mott Insulator, for $2e^{-1/4}<d/l_0<1.86$, one
has a RSB solution without a gap in the conductivity but zero
compressibility, the Mott Glass and finally, for $d/l_0>1.86$, there
is a finite compressibility and no gap in the conductivity, the
Anderson Insulator.  In other words one recovers the solution of
Ref.~\onlinecite{giamarchi_columnar_variat} not as soon as $d/l_0>2e^{-1/4}$ as we would
expect from extrapolations of the perturbative $d=1$ renormalization
group calculations, but only at the higher value $d/l_0> 1.86$.
This is due to the formation of an intermediate phase, which is both
incompressible but without a gap in the conductivity. This
intermediate phase being an intermediate coupling one, the failure of
the perturbative renormalization group to predict its existence is not
a surprise. In a forthcoming section, we will discuss its
properties in detail.  The phase diagram as a function of $d/l_0$ is
represented on Fig.~\ref{fig:phase_diagram}.

Let us remark that all transitions appear to be first order within the
GVM formalism.

\subsection{Mott Insulator to Mott Glass transition: Functional
Renormalization Group study and classical equivalent}
\label{sec:phasefrg}

In this Section we study the phase model (\ref{eq:quantumd}) in
arbitrary dimension $d$ using a renormalization group method
perturbatively controlled in $d=4 -\epsilon$ and small $\hbar$. It
provides useful information on interacting fermions with disorder by
continuation down to $d=1$ (as we do not expect drastic changes in
this model down to $d=1$ for small $\hbar$). We will perform the
analysis in the notations of the classical equivalent model, but also
give some conclusions in terms of the parameters of the quantum model,
via the relations (\ref{eq:corr_class_quant}).

As is well known for classical problems such as manifolds in random
media \cite{fisher_frg_1,fisher_functional_rg,balents_frg_largen}, the
Functional Renormalization Group (FRG) provides an alternative to the
Gaussian Variational Method. The FRG treats accurately the non
linearities and does not use replica symmetry breaking. When the two
methods are supposed to be exact and can be compared they do agree, as
found for the random manifold problem \cite{balents_landscape} for $N
\to \infty$, and generally give consistent physics \cite{giamarchi_vortex_long}.
Here we will use the FRG as a check of the correctness of
the prediction by the GVM of the Mott Glass phase, and as a way to
obtain additional detailed information on the Mott Insulator to Mott
Glass transition.

Although this is not the route we follow here one can apply the FRG
method directly to the model
(\ref{eq:quantumd})--(\ref{eq:acclassic}).
This amounts to
generalize to correlated disorder the study of
\cite{emig_commensurable_frg,emig_rough,emig_commdisorder_long} made
for the case of uncorrelated disorder. It shows that a commensurate
potential becomes relevant and may lead to a description of the
transition between the gapless Anderson Glass to a gapped phase. But
this route would not give us the information we want about the nature
of the gapped phase, i.e it cannot distinguish between the Mott Glass
and a Mott Insulator.  Indeed, the approach of
\cite{emig_commensurable_frg,emig_rough,emig_commdisorder_long} fails
to describe the phase where the commensurate potential is relevant
since the FRG then flows to strong coupling.

Thus, in order to test for the existence of the Mott Glass we consider
from the start a situation where the commensurate potential is
relevant and replace the full
model~(\ref{eq:quantumd})--(\ref{eq:acclassic})
 by an effective model in which the
sine Gordon term is replaced by a quadratic mass term:
\begin{equation} \label{eq:effec}
V_p(\phi) \to \frac{m^2}2 \phi^2.
\end{equation}
This should be a reasonable approximation when the sine Gordon term is
relevant, and is in the spirit of the Self Consistent Harmonic
Approximation (SCHA).  Our approximation amounts to neglecting some
soliton excitations by giving them large energy and to neglecting the
renormalization of the gap by disorder. This simplified model has the
merit to be amenable to a {\it perturbatively} controlled study in
$d=4-\epsilon$. We will show that it does exhibit a phase transition
at $T_{cl}=0$ ($\hbar =0$) which survives at $T_{cl}>0$ ($\hbar >0$)
and can be identified with the Mott Insulator to Mott Glass
transition. This model thus allow to study the formation of the Mott
Glass.

We have obtained the FRG equations for the effective model
(\ref{eq:effec}).  They can be derived by integrated out iteratively
short wavelength modes, extending
\cite{fisher_functional_rg,balents_frg_largen,balents_loc,chauve_mbog_short,chauve_thesis}.
They are obtained in terms of the running dimensionless disorder
$\tilde{\Delta}(\phi,l) = - \tilde{R}^{\prime \prime}(\phi,l)$, the
running dimensionless temperature $\tilde{T}_l \sim T$, both defined
in the Appendix~\ref{ap:frg_approach}, and the tilt modulus
$c_{44}(l)$. These RG equations read:
\begin{eqnarray}
\partial_l \tilde{\Delta}(\phi) &=& \epsilon \tilde{\Delta}(\phi) +
\tilde{T}_l \tilde{\Delta}^{\prime \prime}(\phi) \nonumber \\ && -f(l)
(\tilde{\Delta}^{\prime
\prime}(\phi)(\tilde{\Delta}(0)-\tilde{\Delta}(\phi))-{\tilde{\Delta}^\prime(\phi)}^2)
\\ \partial_l c_{44} &=& -f(l) \tilde{\Delta}^{\prime \prime}(0)
c_{44}
\end{eqnarray}
with $\epsilon=4-d$, setting $c=1$ (as $c$ is not renormalized) and
\begin{equation}
f(l)=\frac{1}{(1+\mu e^{2l})^2}
\end{equation}
comes from the integration of the high momentum modes. Here
$\mu=(m\alpha)^2$.  These FRG equations are analyzed in
Appendix~\ref{ap:frg_approach}.  We describe here only the main
results.

At $T=0$ (i.e $\hbar \to 0$ for the quantum problem) we find that
there is a phase transition. One can measure the strength of the bare
disorder using the Larkin length $R_c \sim
(1/\tilde{\Delta}_{l=0}''(\phi=0))^{1/(4-d)}$ of the problem without
commensurate potential (corresponding to the localization length for
the $d=1$ fermion problem). Then we find that for a given $m$, there
is a transition at a critical disorder strength
\begin{eqnarray}
R_c \equiv
\left(\frac1{\tilde{\Delta}_{l=0}''(\phi=0)}\right)^{1/(4-d)} = C
\frac{1}{m}
\end{eqnarray}
where $C$ is a constant. The two phases are: (i) For strong disorder
$R_c < C \frac{1}{m}$ we find that
$\Delta_2(l)=-\tilde{\Delta}^{\prime \prime}(0,l)$, the fourth
derivative of the renormalized disorder correlator $R_l(u)$, becomes
infinite at a finite scale $R_c^*(m)$, i.e the disorder correlator
becomes non analytic and develops a cusp singularity at a scale
$R_c^*(m)$. For the problem in the absence of a mass the cusp
generation at the Larkin length $R_c^*(m=0) = R_c$ is well known to be
associated to the existence of many metastable states beyond $R_c$.
This cusp generation is associated with the apparition of the
transverse Meissner effect in vortex lattices pinned by columnar
disorder \cite{balents_loc} (and to the appearance of RSB in the GVM
treatment\cite{giamarchi_columnar_variat}). As will be discussed below
this phase corresponds to the {\it Mott Glass}; (ii) For weak disorder
$R_c > C \frac{1}{m}$ the flow is cut by the presence of the mass
before a cusp can be generated. This phase does not exhibit metastable
states and corresponds to the {\it Mott Insulator}.
\begin{figure}
\centerline{\includegraphics[width=\figwidth]{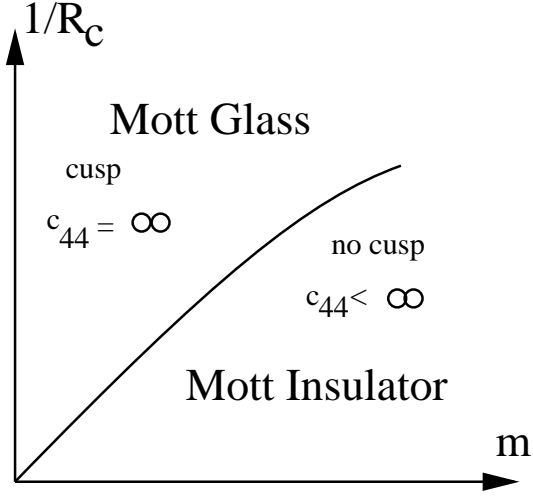}}
\caption{\label{fig:frg1} Phase diagram of the effective model from
the FRG, at $T=0$ ($\hbar =0$). $R_c \sim (1/W)^{1/(4-d)}$ is the
Larkin length (localization length) and parameterizes disorder
strength. These phases and transition survive at $T>0$ ($\hbar >0$).}
\end{figure}

Since the mass can be chosen arbitrarily small the study is thus
perturbative in disorder in $d=4-\epsilon$ for the model
(\ref{eq:effec}).  It is interesting to note that this $T=0$
transition exists both for correlated and uncorrelated
disorder. However, this transition is stable to finite temperature
only for correlated disorder. Indeed for point like disorder the
temperature rounds the cusp
\cite{chauve_creep_short,chauve_creep_long} which implies that there
can exist no sharp distinction between the two phases at finite
temperature. In addition the quadratic part of the Hamiltonian is not
renormalized by disorder, and thus even at $T=0$ there cannot be any
signature of the transition on two point correlation functions of
$\phi$. Thus it is possible that the transition observed in
Ref.~\onlinecite{bouchaud_commdisorder_variational} is an artefact of
the method used. On the contrary for correlated disorder, there is a
genuine transition and because of the lack of rotational invariance
(in $(x,\tau)$), the existence of the cusp and the transition directly
affects for correlated disorder, two point correlation functions.

The FRG gives immediate information on the renormalized tilt modulus
(see (\ref{eq:corr_class_quant}) $c_{44}^R = c_{44}(+\infty)$. Since
this is also the coefficient of $\omega^2$ in the Green function
$\langle \phi \phi \rangle $ one can infer from that that if $c_{44}$
is finite the Green function is likely to remain analytic and thus
{\it that there is a gap in the conductivity}. If $c_{44}$ becomes
infinite then the Green function is not analytic and no gap should
exist in the conductivity. The FRG gives that:
\begin{equation}
\frac{c_{44}(+\infty)}{c_{44}(0)}=\frac{\left(\frac {R_c}
a\right)^\epsilon -1}{\left( \frac {R_c} a \right)^\epsilon
-\left(\frac{R_c^*(\mu)}a \right)^\epsilon}
\end{equation}
and thus we find that  the phase (i) above which corresponds to a
cusp ($c_{44}^R = + \infty$) can be identified with the Mott Glass,
while the phase (ii) above which corresponds to no cusp ( $c_{44}^R <
+\infty$) can be identified with the Mott Insulator.  The gap
itself can be estimated as:
\begin{equation}
\Delta = m/\sqrt{c_{44}^R} \sim (R_c - \frac{C}{m})
\end{equation}
thus it vanishes linearly $\Delta \sim (R_c - \frac{C}{m})$ at the MI
to MG transition. Correlation functions are also estimated in the
Appendix~\ref{ap:frg_approach}.

One of the most crucial test is to show that the above MI-MG
transition survives to quantum fluctuations $\hbar >0$ (thermal
fluctuations $T>0$ for the $d+1$ classical model). This is the case
and the calculation is detailed in Appendix~\ref{ap:frg_approach}. There is no doubt
that the MI survives to quantum fluctuations, however it was less
obvious that the MG would survive. Indeed the cusp is rounded by the
effective temperature variable $\tilde{T}_l \sim T/\sqrt{c_{44}(l)}$.
However the key point is that $c_{44}(l)$ becomes very large as soon
as the second derivative $\Delta_2$ grows, and as a consequence the
effective temperature $\tilde{T}$ renormalizes to exactly zero {\it at
a finite scale} (as it does in the absence of a mass). Thus the Mott
Glass survives at finite temperature. A similar phenomenon was also
found recently in the dynamics of classical periodic systems with
correlated disorder \cite{chauve_mbog_short}.

To conclude this Section, the FRG shows, within a $d=4-\epsilon$
analysis of the effective model with mass, that a transition exists at
large $K$ in the quantum problem (and at low temperature in the
equivalent classical problem) between a MI phase at weak disorder with
analytic Green function, no metastable states and a gap in the
conductivity and a Mott Glass phase with metastable states, no gap in
the conductivity at stronger disorder.  It allows to predict that the
conductivity gap should close linearly at the transition (at least in
the limit of small $K \to 0$).

\section{Physical properties, results in $d=1$ and extensions to higher dimensions} \label{sec:physical_prop}

\subsection{$d=1$}

\subsubsection{Compressibility, density of states and correlation
functions}\label{sec:thermodynamics}

In this section, we define and calculate equilibrium thermodynamic
quantities such as the compressibility or the phason density of states
of the system. One of the most striking difference
between the Anderson Insulator (AI) and the Mott Insulator (MI) is
that the former is compressible whereas the latter is
incompressible. The compressibility is given, in any dimension, by
\begin{widetext}
\begin{equation} \label{eq:density}
\chi(q,\omega_n)=\frac {1}{\hbar}\int d^dx \int_0^{\beta \hbar} d\tau
e^{- i (qx -\omega_n \tau)} \overline{\langle T_{\tau}
(n(x,\tau)-\langle n(x,\tau) \rangle)(n(0,0)-\langle n(0,0) \rangle)
\rangle}
\end{equation}
\end{widetext}
where $n$ is the density. This leads to the average static
compressibility $\chi_s=\lim_{q \to 0}(\lim_{\omega \to 0}
\chi(q,\omega))$.  In $d=1$, using the bosonic expression for the
density (\ref{eq:density}) leads to
\begin{equation}\label{replica_compressiblility}
\chi_s=\lim_{q\to 0} \lim_{\omega \to 0} q^2 G_c(q,\omega)
\end{equation}
where
\begin{eqnarray}
G_c(q,\omega) = \overline{\langle \phi_{q,\omega}
\phi_{-q,-\omega}\rangle - \langle \phi_{q,\omega} \rangle\langle
\phi_{-q,-\omega}\rangle}
\end{eqnarray}

Another thermodynamic quantity of interest is the phason density of states:
\begin{equation} \label{eq:denscdw}
\rho(\omega)=-\frac{\hbar}{\pi v} \Im [ i \omega_n \tilde{G}(x,x, i
\omega_n)]\mid_{ i \omega_n \to \omega+ i 0_+}.
\end{equation}
In the $\overline{K} \to 0$ limit, (\ref{eq:denscdw}) describes the phason
density of states of a classical charge density wave (CDW) pinned
by \emph{both} the commensurate and the random
potential\cite{fukuyama_disorder+commensurate,vinokur_spinchain_random}.
Using the equation (\ref{eq:saddle_point_RS_Gc}), one
obtains the following expression for the density of states $\rho$:
\begin{equation}
  \label{eq:dos_selfenergy} \rho(\omega)=-\frac K 2 \Im \left[\frac {
i \omega_n}{\sqrt{\omega_n^2+\pi \overline{K}(m^2+
I(\omega_n))}}\right]_{ i \omega_n \to \omega +i0_+}
\end{equation}

Various correlation functions can also be computed. In particular the
on-site (CDW) and bond (BOW) charge density. For spinless fermions
they read
\begin{eqnarray}
\chi_{CDW} &=& \langle (c^\dagger_i c_i) (c^\dagger_j c_j)\rangle \\
\chi_{BOW} &=& \langle (c^\dagger_{i+1} c_i + {\rm h.c.})
(c^\dagger_{j+1} c_j + {\rm h.c.})\rangle
\end{eqnarray}
In the boson representation the $2k_F$ part of these correlation
function are related to the $\cos(2\phi)$ and $\sin(2\phi)$
correlation functions
\begin{eqnarray}
\chi_{CDW} &\propto& (-1)^x \overline{\langle
\cos(2\phi(x))\cos(2\phi(0))\rangle} = K_\parallel(x) \\ \chi_{BOW}
&\propto& (-1)^x \overline{\langle
\sin(2\phi(x))\sin(2\phi(0))\rangle} = K_\perp(x)
\end{eqnarray}

Let us now compute these various quantities using the results of the
variational method presented in section~\ref{sec:phasevariational} for
each of the three phases.

\paragraph{Mott Insulator}

It corresponds to the Replica Symmetric (RS) phase obtained at weak
disorder ($d/l_0 <2e^{-1/4}$).  Because of the non--zero
$m$ the whole MI phase is thus incompressible (see
equation (\ref{eq:saddle_point_RS_Gc})). The MI phase is
thus the direct continuation of the non-disordered commensurate
one. In the MI phase, the disorder is too weak to be able to overcome
the gap.

\begin{figure}
\centerline{\includegraphics[width=\figwidth]{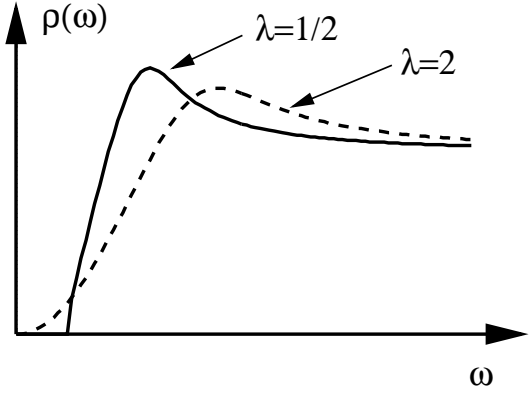}}
\caption{\label{fig:density_of_states} The behavior of the density of
states as a function of $\omega$ for $\lambda=1/2,1,2$. For
$\lambda=1,1/2$ there is a gap in the density of states for
$\omega<\omega_c(\lambda)$. $\omega_c$ decreases with increasing
$\lambda$. For $\lambda=2$, the gap disappears and the density of
states behaves as $\rho(\omega)\sim \omega^2$, i.e. there is a
pseudo-gap. The pseudogap persists in the whole RSB phase. Note that
the maximum in the density of states decreases as $\omega$ increases
indicating the transfer of spectral weight to low frequencies.}
\end{figure}

In the replica symmetric case, $\rho(\omega)$ can be expressed in
terms of the function $f$ defined by equation (\ref{equation-for-f})
in the form:
\begin{equation}
  \label{eq:dos_rs_f} \rho(\omega)=-\frac K 2 \Im \left(\frac x
{\sqrt{1+f(- i x)-x^2}}\right)=\frac K {2\lambda} x \Im f(- i x)
\end{equation}
Where $x=\omega/\omega^*$ and $\omega^*=v/\xi$. To perform the
analytic continuation in Eq. (\ref{eq:dos_rs_f}),
 we transform (\ref{equation-for-f}) into a cubic equation for $f$
with coefficients depending on $x^2$ and $\lambda$. Although this
transformation adds two spurious solutions that do not satisfy
$f(0)=0$, it proves extremely useful as performing the analytic
continuation amounts to solving the cubic equation for $f$ with $x^2 \to
-x^2$. Eq. (\ref{eq:dos_rs_f}) implies that  the phason density of
states in non-zero only when
$f$ has a non-zero imaginary part.  For $\lambda<2$, $f(-ix)$ is
real for :
\begin{equation} \label{critical_x}
x<x_c=\sqrt{1+\lambda-3\left(\frac \lambda 2 \right)^{2/3}}
\end{equation}
As a consequence, in the MI phase, there is a gap in the phason
 density of states for $\omega<\omega_c=\omega^* x_c$
\begin{eqnarray} \label{eq:thresh}
\rho(\omega) = 0 \qquad,\qquad \omega<\omega_c
\end{eqnarray}
The physical interpretation of such form for the density of states is
obvious: no states are available below the gap. Thus, in the Gaussian
Variational framework, there are no discrete two particle states
(i. e. excitons) below the gap.  For $\omega \to \omega_c+0$ we obtain
$\rho(\omega)\sim (\omega-\omega_c)^{1/2}$. At high frequencies, we
obtain $\rho(\omega) \to K/2$ i.e. the density of state goes to a
constant. A plot of $\rho(\omega)$ is shown on
Fig.~\ref{fig:density_of_states}.

In the replica framework, we have the following general expressions
for $K_\parallel$ and $K_\perp$ (see equations
(\ref{eq:correlations_sinus}) and (\ref{eq:correlations_cosinus})):
\begin{eqnarray}
K_\parallel=e^{- \hbar 2 \tilde{G}(0)}\cosh \left( \hbar
2\tilde{G}(x)\right) \\ K_\perp=e^{- \hbar 2\tilde{G}(0)}\sinh \left(
\hbar 2\tilde{G}(x)\right)
\end{eqnarray}

In the replica symmetric case, for $\hbar \to 0$,
$\overline{K}=K/\hbar$ fixed one finds:
\begin{equation}
\hbar \tilde{G}(x)= \frac{\xi^3}{32 l_0^3} \left( 1 +\frac
{|x|}{\xi}\right) e^{-\frac {|x|}{\xi}}
\end{equation}
The resulting correlation functions are:
\begin{eqnarray}
K_\parallel(x)=e^{-\frac{\xi^3}{16 l_0^3}} \cosh \left[\frac{\xi^3}{16
l_0^3} \left( 1 +\frac {|x|}{\xi}\right) e^{-\frac {|x|}{\xi}}\right]
\\ K_\perp(x)=e^{-\frac{\xi^3}{16 l_0^3}} \sinh \left[\frac{\xi^3}{16
l_0^3} \left( 1 +\frac {|x|}{\xi}\right) e^{-\frac {|x|}{\xi}}\right]
\end{eqnarray}
For $x \to \infty$, one has: $K_\parallel(x) \to e^{-\frac{\xi^3}{16
l_0^3}}$. This implies that $\langle \cos 2\phi \rangle =
d/\xi$. Therefore, in the Mott insulator, Charge Density Wave order is
still present, but the order is reduced with respect to the pure
system in which one would have $\langle \cos 2\phi \rangle =1$. Even
for $x\to 0$, the CDW order of the pure system is not recovered.
Another interesting property is that some BOW order is also induced at
short distances, although BOW order is not present at long distance.
The presence of BOW order is due to the random phase in $\phi$ induced
by disorder. When the random phase is of order $\pi/2$, this implies
local BOW order. However, the positions at which BOW order is obtained
are not correlated with each other in the system. This explains the
exponential decay of BOW order.

\paragraph{Anderson glass and Mott glass}

For $d/l_0>1.86$ the system is in the AG phase. In this phase using
(\ref{replica_compressiblility}) and the expression for $G_c$ one
finds that the compressibility is identical to the one of the pure
system $ \chi_s=\frac{\overline{K}}{\pi^2 u}$. Such result is due to
the fact that the Gaussian Variational Method does not take into
account the renormalization of $K$ by disorder. Nevertheless, the
Replica Variational Approximation gives correctly a non-zero
compressibility for an Anderson glass. We stress that these results
are valid independently of the presence and absence
\cite{giamarchi_columnar_variat} of the commensurate potential.

In the intermediate MG phase, with both RSB and a gap that is obtained
for $2e^{-1/4}<d/l_0<1.86$, $m \ne 0$, we obtain a zero
compressibility. One would therefore be tempted to associate this
phase with a Mott Insulator. However, the forthcoming calculation of
the conductivity in Sec. \ref{sec:transport} will show that this
intermediate phase is \emph{not} a Mott Insulator.

In the replica symmetry breaking case, the formulas
(\ref{eq:dos_selfenergy}) and (\ref{eq:dos_rs_f}) remain
valid. However, the function $f$ that must be used in
eq. (\ref{eq:dos_rs_f}) corresponds to $\lambda=2$ in equation
(\ref{equation-for-f}). This means that as long as there is a RSB
solution of the variational equations, there is a pseudogap
 in the phason density of states.  The behavior of the density of
states as a function of $\omega$ is represented on figure
\ref{fig:density_of_states}.

In the case with broken replica symmetry, one has:
\begin{eqnarray}
K_\parallel(x)=e^{ -2 \hbar \langle G(0) \rangle}\cosh \left( 2 \hbar
\langle G(x) \rangle \right) \\
K_\perp(x)=e^{-2 \hbar \langle G(0)
\rangle}\sinh \left( 2 \hbar \langle G(x) \rangle \right)
\end{eqnarray}
Where we have taken into account the fact that as $ \hbar \to 0$,
$\hbar G_c(x) \to 0$, and $\langle G(x) \rangle=\int_0^1 du G(x,u)$.
Using the one step expressions, we obtain:
\begin{widetext}
\begin{equation}
\hbar \langle G(x) \rangle= \frac{e^\varphi}{\mu^3(\varphi)}\left[\left(1
+ \frac{\mu |x|}{2 l_0}\right) e^{-\frac{\mu |x|}{2 l_0}}\right]+
\frac{1-e^\varphi}{2(1-\mu^2(\varphi))}\left[\frac{e^{-\frac{\mu |x|}{2
l_0}}}{\mu}-e^{-\frac{ |x|}{2 l_0}}\right]
\end{equation}
\end{widetext}
We see that in the Mott Glass phase  the CDW order is still
present in analogy with the Mott Insulator. Such behavior is in
agreement with the predictions from atomic limit of
Sec. \ref{sec:phaseatomic}.
This time, $\langle \cos 2 \phi
\rangle=\exp[-e^\phi/(2\mu^3(\phi))]$. When the system becomes an
Anderson Insulator, $\langle \cos 2 \phi\rangle=0$ which seems to
indicate a first order transition. Such first order transition is
likely to be only an artifact of the variational approach. Some
subdominant BOW correlations are also present in the system. They
decay exponentially with $x$ and since $\mu(\phi)<1$
the correlation length of BOW and CDW fluctuations
is  $2l_0/\mu(\varphi)$. It is interesting to note that at the Mott
Insulator-Mott Glass transition, the correlation length is
continuous. However, there could be a slope discontinuity which would
be characteristic of a second order phase transition.

\subsubsection{Transport properties}\label{sec:transport}

To differentiate between a Mott and an Anderson glass, a crucial
physical quantity is the ac conductivity. In the Mott insulator, the
ac conductivity is zero for frequencies smaller than the gap whereas
in the Anderson glass the ac conductivity behaves as
$\sigma(\omega)=\omega^2 (\ln \omega)^2$ in one
dimension\cite{abrikosov_rhyzkin,berezinskii_conductivity_log}.
Within the GVM, in order to compute the conductivity it is sufficient
to know $m$, $\Sigma_1$ and the analytical continuation of
$I(\omega_n)$ to real frequencies.  Using the Kubo formula, it is
straightforward to show \cite{giamarchi_columnar_variat} that:
\begin{equation} \label{conductivite-sigma}
\sigma(\omega)=\frac{v\overline{K}}{\pi}\frac{- i \omega}{\pi
\overline{K}(m^2+I(- i \omega))-\omega^2}
\end{equation}
where $I( i \omega)$ represents the analytic continuation of
$I(\omega_n)$ to real frequencies.  Introducing the function $f$,
defined in (\ref{equation-for-f}) one has:
\begin{equation}\label{conductivite-f}
\sigma(\omega)=\frac{ v \overline{K} }{\pi \omega^*}\frac{ i x}{(1+f(-
i x)-x^2)}
\end{equation}
where $x=\omega/\omega^*$. Similarly to the density of states, the
behavior of the conductivity is therefore controlled by $\lambda=\frac
1 4 (\frac \xi l_0)^3$.  One can explicitly check that
(\ref{conductivite-sigma}) satisfies the sum rule
\begin{equation}
\int_0^\infty d\omega \sigma(\omega) = \frac{v \overline{K}}\pi
\end{equation}

\paragraph{Mott insulator}

\begin{figure}
\centerline{\includegraphics[width=\figwidth]{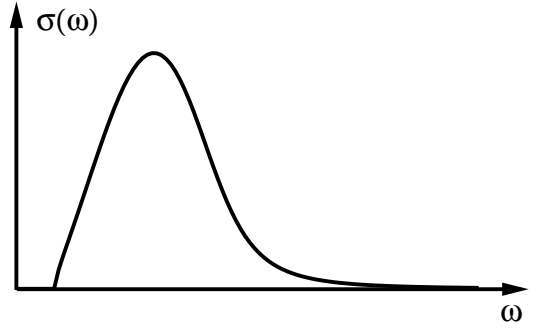}}
\caption{\label{conductivity_lambda=1} The real part of the frequency
dependent conductivity in the MI for $\lambda=1$ as a function of
frequency. }
\end{figure}

\begin{figure}
\centerline{\includegraphics[width=\figwidth]{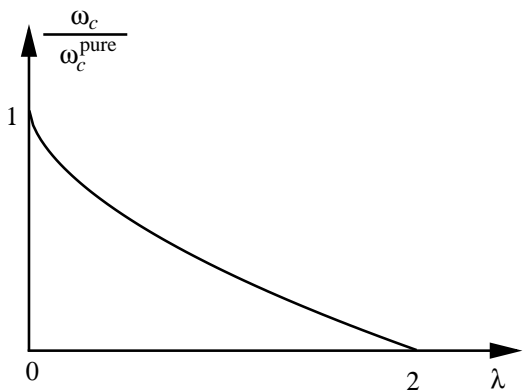}}
\caption{\label{conductivity_gap_lambda} The variation of the gap in
the frequency dependent conductivity as a function of $\lambda$. The
gap in the conductivity goes to zero linearly for $\lambda \to 2$.}
\end{figure}

Let us begin with the conductivity for $\lambda<2$ i.e. in the Mott
Insulator. It is easily seen that in order to obtain a non-zero real
part of the conductivity one must have $\Im f( i x) \ne 0$.  As a
consequence, the real part of the frequency dependent conductivity is
zero for $\omega < \omega_c$ where $\omega_c$ is the threshold below
which the two particle density of states is zero (see
(\ref{eq:thresh}).  Physically, this means that there are no available
two particle excitations to absorb energy if $\omega<\omega_c$ i.e. at
energies below the Mott gap.  For $x>x_c$ ($\omega>\omega_c$), the
analytic continuation of $f$ to imaginary $x$ has a non-zero imaginary
part that leads to a non-zero real part of the frequency dependent
conductivity.  For $x$ close to the threshold,
\begin{equation}
  \label{eq:imaginary_f_above_threshold} \Im f(x)=\frac
{2}{\sqrt{3}}\left(\frac{\lambda} 2\right)^{1/3} \sqrt{x^2-x_c^2}
\end{equation}
As a consequence, for $\omega>\omega_c$ and close to the threshold,
the real part of the conductivity behaves as $\Re \sigma(\omega) \sim
(\omega-\omega_c)^{1/2}$ i.e. it is controlled by the available
two-particle density of states.  At large frequency, it can be shown
that $\Re \sigma(\omega)\sim \frac{\lambda}{x^4}$. This behavior can
be recovered by a simple perturbative calculation in disorder
strength.  Obviously, the conductivity shows a maximum at a frequency
$\omega_m=\omega^*x_m(\lambda)$.  The typical behavior of the real
part of the conductivity for $\lambda=1$ is represented on
Figure~\ref{conductivity_lambda=1}.
The behavior of the threshold frequency $\omega_c$ a function of
$\lambda$ is represented on Figure~\ref{conductivity_gap_lambda}.

\paragraph{Mott glass and Anderson glass}

\begin{figure}
\centerline{\includegraphics[width=\figwidth]{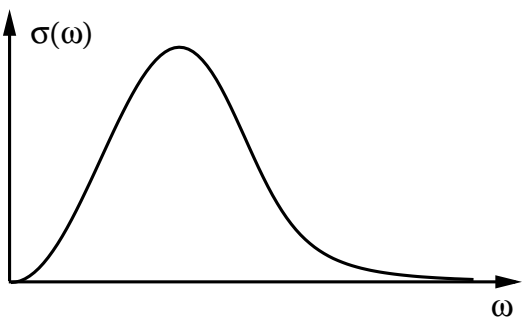}}
\caption{\label{conductivity_lambda=2} The real part of the frequency
dependent conductivity in the MG and AG for $\lambda=2$ as a function
of frequency. For small $\omega$, $\sigma(\omega)\sim \omega^2$. }
\end{figure}

For $\lambda \to 2$ the gap goes to zero as $2-\lambda$.  Quite
remarkably, for $\lambda=2$, there is \emph{no} gap in the real part
of the conductivity although the system is \emph{still}
incompressible. The real part of the conductivity goes to zero as
$\omega \to 0$ as $\Re \sigma(\omega)\sim \omega^2$ . The behavior of
the conductivity for $\lambda=2$ is represented on
Figure~\ref{conductivity_lambda=2}.
As for $\lambda<2$ when $x \to \infty$, the real part of the
conductivity decreases as $\frac \lambda {x^4}$. In fact, this form of
the conductivity is the one that is obtained in the Anderson glass
phase in the \emph{absence} of any commensurate
potential~\cite{giamarchi_columnar_variat}.  Moreover, in the GVM
framework, the Anderson glass is a RSB
phase~\cite{giamarchi_columnar_variat}.  It can be easily seen that
for all $\lambda \ge 2$, i.e. in all the RSB phases, the scaled
conductivity is equal to the one of the Anderson glass. The
conductivity in the MG and AG phases is thus also the one shown on
Fig.~\ref{conductivity_lambda=2}. This remarkable pinning of the
scaled conductivity at $\lambda=2$ is a consequence of the marginality
condition.

\subsubsection{General Phase diagram in $d=1$}

We have thus identified generically three phases for a disordered
commensurate system. The bosonization representation being quite
general in $d=1$ this also applied to bosons or spin chains.

All the previous results having been obtained in the limit where $K$
is small, an important question is the range of stability of these
three phases.  Although, in principle the variational method could
help answering this questions, we do not attempt this complicated
calculation here, and instead give physical arguments.

It is clear that repulsive enough and finite range interactions are
needed for the existence of the MG. A general argument is given in the
following section.  We note here that the case of infinite range
(Coulomb) interaction is a (rather peculiar) example of MG.  Indeed the one dimensional
Wigner Crystal \cite{schulz_wigner_1d} has a compressibility $\chi_s =
\lim_{q\to 0} \frac{q^2}{q^2 \log(1/q)}=0$, nevertheless it has only a
pseudogap in the conductivity\cite{maurey_wigner,shklovskii_conductivity_coulomb}
$\sigma(\omega)\sim \omega^\alpha$.  One can also show that in a
non-interacting system, the compressibility gap is equal to the gap
for single particle excitations.
In particular, this means that the intermediate phase cannot exist for
$K=1$.  This result is in agreement with the Self Consistent Born
Approximation calculation of Mori and Fukuyama\cite{mori_scba} for the
non interacting case , which do not show any intermediate phase. Thus
it can only exist for $K\le K_c <1$.

\begin{figure}
\centerline{\includegraphics[width=\figwidth]{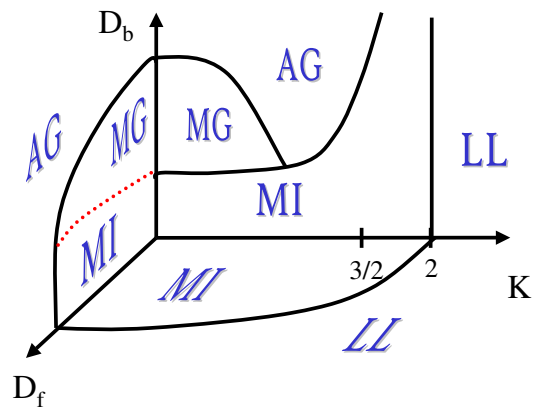}}
\caption{\label{fig:phase_diagram_forback} Phase diagram of a one
dimensional system with both forward and backward scattering random
potential. The dashed lines correspond to phase boundaries between the
Mott glass (MG) and the Mott Insulator (MI), the Anderson Insulator
(AI) and the Luttinger liquid (LL) phase. The separation between the
MG and the MI phase in the presence of forward scattering disorder is
drawn with question marks since we do not know how forward scattering
affects the competition of MI and MG phases.}
\end{figure}

Let us now give a schematic phase phase diagram, which summarizes the
effects of both the backward and forward scattering in one
dimension. As shown in section \ref{sec:forward-periodic} forward
scattering can also lead to gap closure.  The phase diagram, as
function of the LL parameter $K$ and the strength of the forward $D_f$
and backward $D_b$ scattering is represented in
Fig.~\ref{fig:phase_diagram_forback}

\subsection{General arguments and higher dimensions $d>1$} \label{sec:fermionshigh}

\subsubsection{Interacting fermionic systems: excitonic argument} \label{sec:excitonic}

\begin{figure}
\centerline{\includegraphics[width=\figwidth]{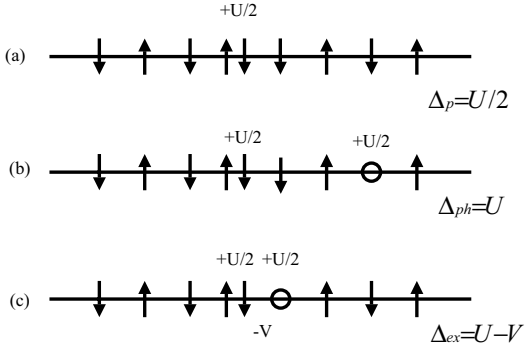}}
\caption{\label{fig:excitonic} Possible excitations in the atomic limit. Drawings are made for a chain
for clarity but the arguments are valid in arbitrary dimension $d$. (a) Energy cost to add one particle.
(b) Generic Particle hole excitation. (c) Exciton, where the particle and the hole are on neighboring sites.
In the presence of disorder the gap for excitonic excitations will close first leading to the absence
of a gap in the optical conductivity, but still to an incompressible system (see text), leading to the Mott Glass.}
\end{figure}

The physics leading to the MG phase is quite general and persists in
higher dimension as well, as can be understood through a physical
argument.  Let us consider the atomic limit, where the hopping is
zero. One can compute in this limit the gaps to create both single
particle and particle-hole excitations (see fig.~\ref{fig:excitonic})
Let us consider for example fermions with spin with both an onsite
repulsion $U$ and a nearest neighbor repulsion $V$, with one particle
per site. Such a system is described by
\begin{eqnarray}
H &=& U\sum_i n_{i\uparrow}n_{i\downarrow} + V \sum_{\langle
i,j\rangle} n_i n_j \nonumber \\ & & + \sum_i W_i n_i - \mu \sum_i n_i
\end{eqnarray}
where $n_i = n_{i\uparrow} + n_{i\downarrow}$, and $W_i$ is the
disorder potential.  The energy to add $E_{+1}$ or remove $E_{-1}$ a
particle at/from site $i$ are
\begin{eqnarray} \label{eq:single}
E_{+1,i} &=& E_{\text{gs}} + U + z V - \mu + W_i\\ E_{-1,i} &=&
E_{\text{gs}} - z V + \mu - W_i
\end{eqnarray}
and $E_{\text{gs}}$ the energy of the ground state of the system with
one particle per site.  If one considers a particle hole excitation
where the particle moves from the site $i$ to site $j$, the energy
cost is $E_{+1,j} - E_{-1,i}$ if $i,j$ are not nearest neighbors.  On
the other hand if the particles are nearest neighbors (excitonic
excitation), this costs
\begin{eqnarray} \label{eq:parthole}
\Delta_{\text{ph},ij} = U - V + W_j - W_i
\end{eqnarray}
For the pure case, one thus sees from (\ref{eq:single}) and
(\ref{eq:parthole}) that the gap for creating a single particle
excitation is larger than for particle hole
\begin{eqnarray}
\Delta_{\text{p}} &=& \frac{U}2 \\ \Delta_{\text{ex}} &=& U - V
\end{eqnarray}
this is the well known excitonic binding that occurs in systems with a
gap.

In the presence of disorder one can minimize the single particle gap
by choosing the site where the disorder potential is minimum, giving
\begin{eqnarray}
\Delta_{\text{p}} &=& \frac{U}2 + \frac{\min(W_i) - \max{W_i}}2
\end{eqnarray}
where we choose $\overline{W_i} = 0$ for convenience.  On the other
hand the minimal particle hole interaction corresponds to choosing the
nearest neighbor pair $\langle i,j \rangle$ for which the difference
in disorder potential is minimal
\begin{eqnarray}
\Delta_{\text{ex}} = U - V - \min_{\langle i,j \rangle} |W_j - W_i|
\end{eqnarray}
For an uncorrelated bounded disorder one has
\begin{eqnarray}
\min(W_i) \sim -W \\ \min(W_j - W_i) \sim -2W
\end{eqnarray}
Thus, in presence of a nearest neighbor interaction $V$, the particle
hole gap closes faster, at $W_c = (U-V)/2$, when disorder increases,
than the single particle one. For an homogeneous system this would
simply signals an instability of the ground state. For the disordered
one this need not be so, since only a fraction of the sites have their
gap closing. Thus, in the presence of a small kinetic energy the
conductivity gap would close near this point, the compressibility
remaining zero.  Within this zero kinetic energy model one thus
already finds three phases.  The phase for which the particle hole gap
has closed for some sites but the single particle gap is still finite
can of course be identified with the Mott glass.

Thus the physics of the Mott glass, that has been derived for finite
kinetic energy by the methods of the previous sections has its origin
in excitonic effects. This is quite general and does not rely on any
special one dimensional features. One dimension was thus here only a
tool allowing us to perform the calculation.  We thus expect the Mott
glass to be present in arbitrary dimension, and it would be
interesting to check either through numerical calculations or mean
field methods whether one can recover the properties that we have
identified here.  The excitonic argument also shows clearly that some
finite range interaction is needed for the MG to appear. For a simple
Hubbard model both the single particle and particle-hole gap would
close simultaneously (up to the distribution of disorder) and most
likely the MG phase does not exist. In the presence of finite range
interactions the MG glass can be stabilized.

A similar construction can be made for the spinless case, although it
involves longer range (third neighbor) interactions.

According to this physical picture of the MG, the low frequency
behavior of conductivity is dominated by excitons (involving
neighboring sites). This is at variance from the AG where the particle
and the hole are created on distant sites. This has consequences on
the precise low frequency form of the conductivity such as logarithmic
corrections. In addition since the excitons are neutral objects,
although they can participate to the optical absorption, they need to
be broken to give a d.c. current. One can thus give a naive estimate
of the conductivity in the MG
\begin{eqnarray}
\sigma \sim n_{ex} e^{-\frac{V}T}
\end{eqnarray}
where $n_{ex}$ is the number of excitons in the ground state and $V$
is the typical excitonic binding energy, which depends only weakly on
the disorder.

\subsubsection{Consequences for other systems}

The above arguments also directly apply to other systems. In one
dimension the spinless fermions can be mapped to a disordered spin
systems. In that case the commensurate phase can either come from an
antiferromagnetic staggered field, or more reasonably from a
spin-Peierls distortion of the lattice.  Such a perturbation would
force the spin to lock into a singlet state. The disorder would be a
random magnetic field.

\begin{figure}
\centerline{\includegraphics[width=\figwidth]{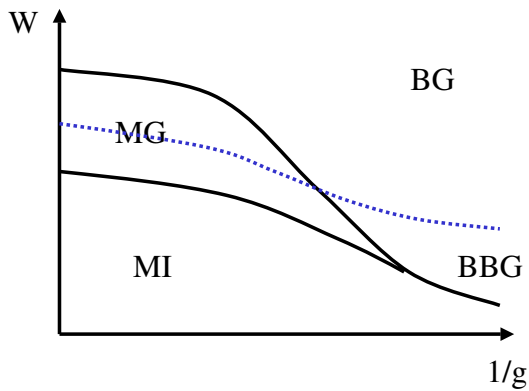}}
\caption{\label{fig:braggbose} Phase diagram of bosons in the limit
$T\to 0$, $\hbar \to 0$.  $W$ is the strength of disorder and $g$ the
amplitude of the commensurate potential for a fixed repulsive
interactions.  For weak disorder the boson system should retain
perfect topological order (i.e. no defects such as dislocations).
Whether the line for which topological order is lost (dashed line)
enters the MG phase or not is an open question.  It is represented
here for $d=3$. The incompressible phases are the Mott Insulator (MI) or
Mott Glass (MG). The compressible ones are either the Bragg Bose Glass
with perfect topological order (BBG), or the Bose Glass (BG) if topological order
is lost.}
\end{figure}

Another system of interest is provided by hard core bosons. In one
dimension, one can use exactly the phase Hamiltonian to represent
interacting bosons \cite{haldane_bosons,giamarchi_loc}, but the
excitonic arguments given in Sec.~\ref{sec:excitonic} would also apply
to interacting bosons in higher dimensions as Well.  In that case the
Anderson glass becomes the Bose glass
\cite{giamarchi_loc,fisher_boson_loc}.  For classical systems the
phase diagram is shown on Fig.~\ref{fig:braggbose}. Even in the Bose
glass phase, for weak disorder perfect topological order (stability to
dislocations in the lattice) can persist in $d=3$ resulting in a Bragg
Bose glass phase
\cite{giamarchi_vortex_long,giamarchi_diagphas_prb}. On the other hand
in any dimension the Mott insulator should exhibit perfect topological
order. Thus an interesting and open issue is whether this topological
order also subsist in at least a portion of the Mott glass phase.

\section{Conclusion}\label{sec:conclusion}

In this paper, we have investigated the competition of a random and a
commensurate potential. This question is relevant for host of physical
systems, ranging from one dimensional interacting fermions or bosons,
to classical systems in the presence of correlated disorder.  The
commensurate potential induces an incompressible Mott insulating phase
with a gap in the conductivity. On the other hand, disorder induces a
compressible Anderson insulating phase.  While naive expectations
predict a direct transition between these two phases, we find that if
interactions are repulsive enough, an intermediate phase, the Mott
glass, does exist. Although this phase is incompressible, as a Mott
insulator, it does not have a gap in the optical conductivity, in a
way similar to the Anderson insulator.

To obtain this phase we had to go beyond standard renormalization
group techniques which are perturbative in both the commensurate and
disorder potential.  We therefore have used bosonization associated
with several non perturbative techniques.  The first one is a replica
variational method, that allows for a complete calculation of the
various physical observables such as the conductivity. The second
method is a functional renormalization group, which is perturbative in
$d=4-\epsilon$ dimensions and is well suited to study the transition
from the Mott glass to the Mott insulator, as well as equivalent
classical systems. In addition we have looked at the limit of zero
kinetic energy, both for the bosonized Hamiltonian, and directly on
the fermion problem (both for the spinless problem and for the problem
with spin).  The later yields a very general argument in favor of the
existence of the Mott glass in any dimension.  It also shows that the
underlying mechanism for this phase is the creation of low energy
bound states (excitons) coming from the competition between
interactions and disorder. These excitations play no role in the
compressibility but contribute to the optical conductivity.

This phase could be observable in systems close to a metal insulator
transition, such as oxides, provided that one can measure
simultaneously the optical conductivity and the
compressibility. Numerical simulations on disordered boson systems
could be prime candidates to observe this effect. Note that since all
the phase have a finite correlation length, this should be observable
even in moderately small systems.
Many problems remain open. In particular, it would be interesting to
understand in detail the effect of a chemical potential on the
Mott-Glass phase. Another open problem is the effect of temperature on
the Mott Glass phase. Finally, it would be interesting to investigate
the possibility of aging dynamics in the Mott Glass.

\begin{acknowledgments}
We thank R. Bhatt, S. Fujimoto, H. Fukuyama, A. Furusaki, L. Ioffe,
C. Itoi, N. Nagaosa,  Y. Suzumura, C. M. Varma and H. Yoshioka for
discussions. E. O. acknowledges support from NSF under grants DMR
96-14999 and DMR 9976665 (during his stay at Rutgers University where
part of this work was completed) and from Nagoya University.
\end{acknowledgments}

\appendix

\section{Forward scattering disorder and periodic potential}\label{sec:forward-periodic}

In this appendix we examine the effects of the forward scattering
disorder $\eta$ and neglect altogether backward scattering. For
fermions, such an approximation is surely justified when
$3/2<K<2$. Then, the backward component of disorder is irrelevant and
can be neglected. In the other case, $K<3/2$, backward scattering will
be relevant and drive the system into an Anderson glass state.

\subsection{solution of the variational equations} \label{sec:solvar}

The action of the problem is:
\begin{widetext}
\begin{equation}\label{eq:action_forward}
\frac S \hbar = \int dx \int_0^{\beta \hbar} d \tau \left[\frac 1
{2\pi K} \left\{ v(\partial_x \phi)^2 +\frac{(\partial_\tau \phi)^2}
v\right\} - \frac g {\pi \alpha \hbar} \cos 2\phi -\frac{\mu(x)}{\pi
\hbar} \partial_x \phi \right]
\end{equation}
\end{widetext}
which gives after replication and average over disorder
\begin{widetext}
\begin{eqnarray}\label{replicated-action}
S_{\text{rep.}}&= &\sum_a \int dx \int_0^{\beta \hbar} d\tau \left[ \frac
1 {2\pi K}\left\{ v(\partial_x \phi_a)^2 +\frac{(\partial_\tau
\phi_a)^2} v\right\} -\frac { g} { \pi \alpha } \cos 2 \phi_a \right]
\nonumber \\
&- &\frac D {2 (\pi \hbar)^2} \sum_{a,b} \int dx \int_0^{\beta \hbar}
d\tau \int_0^{\beta \hbar} d\tau' \partial_x \phi_a(x,\tau) \partial_x
\phi_b(x,\tau')
\end{eqnarray}
\end{widetext}
We use the GVM ansatz (\ref{eq:variational_action}) with
\begin{equation}
v G^{-1}_{ab}(q,\omega)=\frac {((vq)^2+\omega^2)}{\pi K}\delta_{ab} -
\sigma_{ab}(q,\omega)
\end{equation}
as in the case of the backward scattering disorder.  Using
(\ref{eq:variational_action}) and (\ref{equ:varbase}) the variational
energy $F_{\text{var}}$ for the forward scattering problem is:
\begin{eqnarray} \label{equ:explicit}
F_{\text{var}}&=&\frac 1 {2 \beta} \int \frac {dq}{2\pi} \sum_{n,a}
\frac \hbar {\pi K}(v
q^2+\frac{\omega_n^2}{v})G_{aa}(q,\omega_n)\nonumber \\ &-&\frac 1 {2
\beta} \int \frac{dq}{2\pi} \sum_{a,n} (\ln G)_{aa}(q,\omega_n)
\nonumber \\ &-&\frac{g}{\pi \alpha }\sum_a \exp \left(-2 \hbar
G_{aa}(x=0,\tau=0)\right) \nonumber \\ &-& \frac{D}{2 \pi^2
\hbar}\sum_{a,b} \int \frac{dq}{2\pi} \hbar q^2 G_{ab}(q,\omega_n=0)
\end{eqnarray}
Minimizing (\ref{equ:explicit}) with respect to $G(q,\omega)$ gives
the variational equations
\begin{eqnarray} \label{equ:varfor}
\sigma_{ab}(q,\omega_n)=-\frac{ Dv q^2 \beta}{\pi^2}
\delta_{\omega_n,0}-\frac{4gv}{\pi \alpha} e^{-2 \hbar
G_{aa}(x=0,\tau=0)}\delta_{a,b}
\end{eqnarray}
It is easy to check that the equations (\ref{equ:varfor}) only have
replica symmetric solutions in contrast to the case of backward
scattering.

Using the standard techniques for inversion of matrices in the limit
$n \to 0$ \cite{mezard_variational_replica}, one finds the following
expressions for $G_c=G_{aa}+\sum_{b\neq a}G_{ab}$:
\begin{equation}
G_c(q,\omega_n)=\frac v {\frac \hbar {\pi K}(\omega_n^2+(vq)^2)+ m^2}
\end{equation}
and for $G(q,\omega_n,u)$:
\begin{equation}
G(q,\omega_n,u)=\left(\frac{\pi \overline{K}} \hbar\right)^2\frac
{Dq^2\beta \delta_{\omega_n,0} }{ \pi^2 v^2 (q^2+\xi^{-2})^2}
\end{equation}
where:
\begin{equation}
m^2=\frac{4 g v}{\pi \alpha} e^{-2 \hbar G_{aa}(x=0,\tau=0)}
\end{equation}
and $\xi^2=v^2/(\pi \overline{K} m^2)$. One has:
\begin{equation}
\lim_{\hbar \to 0, \overline{K} \text{fixed}} \hbar
\tilde{G}(0,0)=\frac \pi 2 \frac{D\overline{K}^2}{v^2}\xi
\end{equation}
Leading to the self-consistent equation for $\xi^2$ :
\begin{equation}\label{eq:self_consistent_forward_t>0}
\left(\frac{l_1}{\xi}\right)^2 \exp\left (\frac
{\xi}{l_1}\right)=\left(\frac{l_1}{d}\right)^2
\end{equation}
Where we have defined $l_1^{-1}=\frac{ D \overline{K}^2}{2 v^2}$.

It is straightforward to show that
(\ref{eq:self_consistent_forward_t>0}) has two solutions for $l_1/d
>e/2$ and no solutions otherwise. In the first case, the solution with
$l_1/\xi<1/2$ is a spurious solution, as can be seen by taking the
limit of zero disorder ($l_1 \to \infty$). Physically, the Mott gap is
preserved (but reduced) as long as $l_1/d>e/2$, whereas for
$l_1/d<1/2$, the Mott gap disappears. The transition gapped--gapless
appears first order in the GVM, which is likely to be an artifact of
the variational method. Note that the condition for the transition
$l_1/d \sim e/2$ is in agreement with strong coupling extrapolations of
the perturbative RG treatment.

\begin{figure}
\centerline{\includegraphics[width=\figwidth]{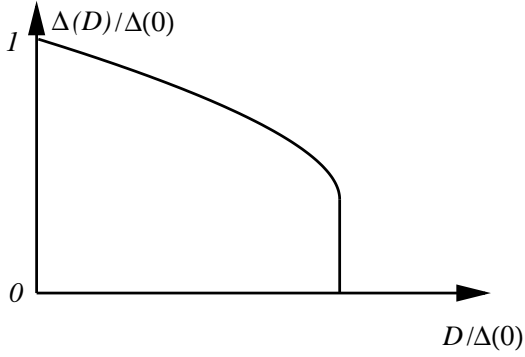}}
\caption{\label{fig:gap_disorder_forward} The reduction of the gap
with forward scattering disorder strength}
\end{figure}

\subsection{Correlation functions}

Another advantage of the GVM is to allow for the calculation of the
correlation functions. Using (\ref{eq:density}) one sees that the
density-density correlation functions is given by
\begin{equation}\label{eq:correlations}
\overline{\langle T_\tau e^{ i 2\phi(x,\tau)} e^{\pm i
2\phi(0,0)}\rangle}=e^{- \hbar \left[2\tilde{G}(0,0) \pm
2\tilde{G}(x,\tau)\right]}
\end{equation}
Which leads to:
\begin{eqnarray}
K_\perp(x,\tau) &=& \overline{\langle T_\tau \sin(2\phi(x,\tau))
\sin(2\phi(0,0))\rangle} \nonumber \\ &=& e^{- \hbar 2\tilde{G}(0,0)}
\sinh \hbar 2\tilde{G}(x,\tau) \label{eq:correlations_sinus} \\
K_\parallel(x,\tau) &=& \overline{\langle T_\tau \cos (2\phi(x,\tau))
\cos (2\phi(0,0))\rangle} \nonumber \\ &=& e^{- \hbar 2\tilde{G}(0,0)}
\cosh \hbar 2\tilde{G}(x,\tau)\label{eq:correlations_cosinus}
\end{eqnarray}
Since $\tilde{G}(x,\tau)=G_c(x,\tau)+G(x)$ and $\lim_{\hbar \to 0}
\hbar G_c(x,\tau)=0$, only the static correlations survive.  It is
straightforward to show that:
\begin{equation}
\lim_{\hbar \to 0, \overline{K} \text{fixed}} \hbar G(x)=\frac 1 2
\left(\frac \xi {l_1} -\frac{|x|}{l_1}\right) e^{ -\frac{|x|} \xi}
\end{equation}
Leading to the following expressions for the correlation functions:
\begin{eqnarray}
K_\perp(x)=e^{-\frac \xi {l_1}} \sinh \left[ \left(\frac \xi {l_1}
-\frac{|x|}{l_1}\right) e^{ -\frac{|x|} \xi}\right] \\
K_\parallel(x)=e^{-\frac \xi {l_1}} \cosh \left[ \left(\frac \xi {l_1}
-\frac{|x|}{l_1}\right) e^{ -\frac{|x|} \xi}\right]
\end{eqnarray}
It is easily seen that for $\xi < \infty$, $\lim_{x \to \infty}
K_\parallel=e^{-\frac \xi {l_1}}$. In the absence of disorder, this
limit would be exactly one. Forward scattering disorder thus leads to
a reduction of the Charge Density Wave long range order.  For $\xi \to
\infty$, one recovers $K_\parallel(x) \sim e^{-\frac{|x|}{l_1}}$, a
result that could have been derived directly.

\section{Saddle point equations} \label{sec:appendix_gvm}

We derive in this appendix the saddle point equations obtained by
minimizing the variational free energy (\ref{equ:varbase}). Using
(\ref{equ:varbase}) and (\ref{eq:pp+cdb_replica}) we get
\begin{widetext}
\begin{eqnarray}\label{eq:var_fe_pp+cdb}
F_{\text{var}}&=&\frac 1 {2 \beta} \int \frac {dq}{2\pi} \sum_{n,a}
\frac \hbar {\pi K}(v q^2+\frac{\omega_n^2}{v})G_{aa}(q,\omega_n) -
\frac 1 {2 \beta} \int \frac{dq}{2\pi} \sum_{a,n} (\ln
G)_{aa}(q,\omega_n) \nonumber \\ & & - \frac{g}{\pi \alpha }\sum_a
\exp \left(-2 \hbar G_{aa}(x=0,\tau=0)\right) - \frac{W}{(\pi \alpha
)^2\hbar }\int_0^{\beta\hbar} d\tau \sum_{a,b} \exp \left[ - 4
\hbar\left(G_{aa}(x=0,\tau=0)-G_{ab}(x=0,\tau)\right)\right]
\end{eqnarray}
\end{widetext}
Varying in (\ref{eq:var_fe_pp+cdb}) with respect to $G$ we get the
following saddle point equations:
\begin{widetext}
\begin{eqnarray}
G_c(q,\omega_n)^{-1} &=& \frac \hbar {\pi K} (v
q^2+\frac{\omega_n^2}{v})+\frac {4g}{\pi \alpha} \exp(- 2 \hbar
G_{aa}(x=0,\tau=0)) \nonumber \\ & & + \frac{ 2W }{\hbar(\pi\alpha
)^2}\int_0^{\beta \hbar}d\tau (1-\cos(\omega_n \tau)) \left[e^{-4
\hbar B_{aa}(x=0,\tau)} + \sum_{b\neq a} e^{-4 \hbar
B_{ab}(x=0,\tau)}\right]
\label{saddle-point-backward-gc} \\
\sigma_{a \neq b}(q,\omega_n)&=&\frac{2 W v }{ \hbar (\pi
\alpha)^2}\int_0^{\beta\hbar} d\tau\cos (\omega_n \tau) \exp \left(-4
\hbar B_{ab}(x=0,\tau) \right)
\label{saddle-point-backward-sigma}
\end{eqnarray}
\end{widetext}
with $G_c=G_{aa}+\sum_{b\neq a} G_{ab}$ and
\begin{equation}
B_{ab}(x,\tau)=G_{aa}(x,\tau)-G_{ab}(x,\tau)
\end{equation}
As was the case for fermions in a random potential
\cite{giamarchi_columnar_variat}, one has: $\frac{dB_{a\neq
b}}{d\tau}=0$ leading to the following simplified expression of the
replica off diagonal self energy:
\begin{equation}
\sigma_{a \neq b}= \frac{2 W v}{(\pi \alpha)^2} \beta \exp (-4 \hbar
B_{a \neq b})\delta_{\omega_n,0}
\end{equation}
We still need to perform the analytical continuation from positive
integer $n$ to $n=0$ in equations (\ref{saddle-point-backward-gc}) and
(\ref{saddle-point-backward-sigma}).  In the GVM this is done assuming
that for $n\to 0$, the $G_{ab}$ become hierarchical matrices. Using
the Parisi parameterization of hierarchical matrices in the $n \to 0$
limit\cite{mezard_variational_replica}.
(\ref{saddle-point-backward-gc}) and
(\ref{saddle-point-backward-sigma}) give the following equations:
\begin{widetext}
\begin{eqnarray}
G_c^{-1}(q,\omega_n) & = & \frac \hbar {\pi K} (v
q^2+\frac{\omega_n^2} v) +\frac{4 g }{\pi \alpha } \exp\left(-2 \hbar
\tilde{G}(x=0,\tau=0)\right) \nonumber \\ & & + \frac{2W}{\hbar(\pi
\alpha )^2}\int_0^{\beta \hbar}d\tau (1-\cos(\omega_n \tau))
\left[\exp(-4\hbar \tilde{B}(x=0,\tau))-\int_0^1 du \exp (-4 \hbar
B(u))\right]\label{saddle-point-backward-appendix} \\
\sigma(q,\omega_n,u) &=& \frac{2 W v }{ (\pi \alpha)^2}\beta \exp
(-\hbar 4 B(u))\delta_{\omega_n,0}
\end{eqnarray}
\end{widetext}
where $u \in [0,1]$ is the Parisi parameter replacing the discrete
replica index $a$.

\section{Solution of RSB equations} \label{sec:appendix_rsb}

\begin{figure}
\centerline{\includegraphics[width=\figwidth]{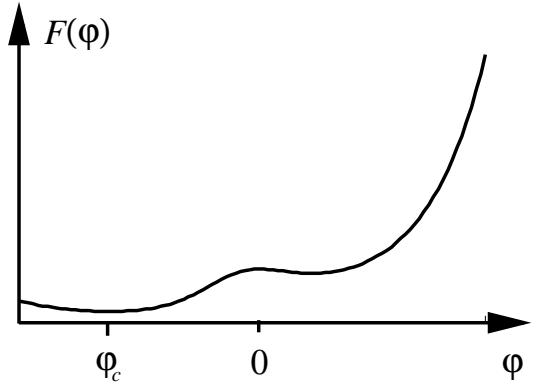}}
\caption{\label{fig:f_of_phi} The graph of $F(\varphi)$. Only the
region
with $\varphi<0$ is
physical.  $F$ has a
minimum for  $\varphi=\varphi_c$.  When $4(l_0/d)^2$ is smaller than
$F(\varphi_c)$, equation (\ref{eq:F-of-phi}) has no solution.}
\end{figure}

We want to solve the RSB saddle point equations
\begin{widetext}
\begin{eqnarray}
v G_c^{-1}(q,\omega_n)&=&\frac 1 {\pi \overline{K}}((v
q)^2+\omega_n^2) + m^2+\Sigma_1(1-\delta_{n,0}) +I(\omega_n) \\
I(\omega_n)&=&\frac{2 W v}{ (\pi \alpha v)^2 \hbar
}\int_0^{\beta\hbar} \left[e^{-4\hbar \tilde{B}(\tau)}-e^{-4\hbar
B(u_c)}\right](1-\cos(\omega_n \tau)) d\tau \\
\Sigma_1&=&u_c(\sigma(u>u_c)-\sigma(u<u_c)) \\ \sigma(u)&=&\frac{2 W
v}{ (\pi \alpha)^2}e^{-\hbar 4 B(u)}\beta \delta_{n,0} \\ m^2=\frac{4
g v}{\pi \alpha}e^{-4 \hbar \tilde G (0)}
\end{eqnarray}
\end{widetext}
Using the inversion formulas for hierarchical
matrices\cite{mezard_variational_replica},
\begin{widetext}
\begin{eqnarray}
\tilde{G}(q,\omega_n=0)-G(q,\omega_n=0,u<u_c) &=&
\frac{1}{G_c^{-1}(q,\omega_n=0)} +\left(1-\frac 1
{u_c}\right)\left(\frac{1}{G_c^{-1}(q,\omega_n=0)+\frac{\Sigma_1} v}
-\frac{1}{G_c^{-1}(q,\omega_n=0)}\right) \\
\tilde{G}(q,\omega_n=0)-G(q,\omega_n=0,u<u_c) &=&
\frac{1}{G_c^{-1}(q,\omega_n=0)+\frac{\Sigma_1} v}
\end{eqnarray}
\end{widetext}
we get:
\begin{widetext}
\begin{eqnarray}
B(u<u_c) &=& \frac{1}{\beta \hbar} \sum_{\omega_n} \int
\frac{dq}{2\pi} G_c(q,\omega_n) + \frac{1}{\beta \hbar} \left(1-\frac
1 {u_c}\right) \int \frac{dq}{2\pi}
\left(\frac{1}{G_c^{-1}(q,\omega_n=0)+\frac{\Sigma_1} v}
-\frac{1}{G_c^{-1}(q,\omega_n=0)}\right) \\ B(u>u_c) &=&
\frac{1}{\beta \hbar} \sum_{\omega_n} \int \frac{dq}{2\pi}
G_c(q,\omega_n) + \frac 1 {\beta \hbar} \int \frac{dq}{2\pi}
\left[\frac{1}{G_c^{-1}(q,\omega_n=0)+\frac{\Sigma_1}{v}}-\frac
{1}{G_c^{-1}(q,\omega_n=0)}\right] .
\end{eqnarray}
\end{widetext}
Since
\begin{equation}
\lim_{\beta \to \infty \atop \hbar \to 0} \frac{1}{\beta \hbar}
\sum_{\omega_n} \int \frac{dq}{2\pi} G_c(q,\omega_n) = 0
\end{equation}
We obtain:
\begin{eqnarray}
\lim_{\beta \to \infty \atop \hbar \to 0} -\hbar B(u>u_c)&=&0 \\
\lim_{\beta \to \infty \atop \hbar \to 0} -4 \hbar B(u<u_c)&=& 2
\frac{\sqrt{\pi \overline{K}}}{ \delta}\left[ \frac 1
{(m^2+\Sigma_1)^{1/2}} - \frac 1 m \right] \nonumber
\end{eqnarray}
Here, we have assumed that when $\beta$ goes to infinity, $u_c$ goes
to zero in such a way that $\beta u_c=\delta$ remains finite.
Therefore
\begin{equation}
\Sigma_1=\frac{2W}{(\pi\alpha)^2} \delta v \left(1-e^{2
\frac{\sqrt{\pi \overline{K}}}{\delta}\left(\frac 1
{(m^2+\Sigma_1)^{1/2}} -\frac 1 m \right)}\right)
\end{equation}
Next, we derive a self-consistent equation for $m$ by taking the
$\hbar\to 0$ limit of the equation:
\begin{equation}
m^2=\frac{4 g v}{\pi \alpha }e^{-2 \hbar\tilde{G}(0)}
\end{equation}
We use first the general inversion formula
\cite{mezard_variational_replica}:
\begin{widetext}
\begin{eqnarray}
\tilde{G}(q,\omega_n)=\frac{1}{G_c^{-1}(q,\omega_n)}\left [ 1-\int_0^1
\frac{du}{u^2}
\frac{[G^{-1}](u)}{G_c^{-1}-[G^{-1}](u)}-\frac{G^{-1}(0)}{G_c^{-1}}\right](q,\omega_n)
\end{eqnarray}
\end{widetext}
In which we have:
\begin{eqnarray}
\lbrack G^{-1}\rbrack (u<u_c)=0 \\ \lbrack G^{-1}\rbrack
(u>u_c)=\frac{-\Sigma_1} v \\ G^{-1}(0)=-\frac{\sigma(u<u_c)}{v}
\end{eqnarray}
so that:
\begin{widetext}
\begin{eqnarray}
\tilde{G}(0,0)=\frac{1}{\beta \hbar} \int \frac{dq}{2\pi}
\sum_{\omega_n} G_c(q,\omega_n) + \frac{1}{\beta \hbar} \int
\frac{dq}{2\pi} \left[ \frac{ v \sigma(u<u_c)}{\left(\frac{\hbar}{\pi
K}(vq)^2 +m^2\right)^2} +\left(1- \frac 1 {u_c}\right) \frac{v
\Sigma_1}{\left(\frac{\hbar}{\pi K}(vq)^2 +m^2\right)
\left(\frac{\hbar}{\pi K}(vq)^2 +m^2 +\Sigma_1 \right)}\right]
\end{eqnarray}
leading to the expression for $m$ in the $\beta \to \infty$ limit:
\begin{eqnarray}
m^2& = &\frac{4 g v}{\pi \alpha}\exp \left[ 2 \frac{\Sigma_1(\pi
\overline{K})^{1/2}}{\delta} \frac{1}{m
(m^2+\Sigma_1)^{1/2}[m+(m^2+\Sigma_1)^{1/2}] } - \frac{ W(\pi
\overline{K})^{1/2}v}{(\pi \alpha)^2 m^3}\exp\left(2 \frac{ \sqrt{\pi
\overline{K}}}{\delta} \left[ \frac 1 {(m^2+\Sigma_1)^{1/2}} -\frac 1
m \right]\right) \right]
\end{eqnarray}
\end{widetext}
A final equation for the breakpoint $u_c$ is needed to close the
system of equations. As was discussed in
Ref. \onlinecite{giamarchi_columnar_variat} the physical choice
corresponds to the so called marginality of the replicon condition
which yields to $I(\omega_n) \propto \mid \omega_n \mid$ and to :
\begin{equation}\label{marginality_condition}
\frac{4 W (\pi \overline{K})^{1/2} v}{(\pi \alpha)^2
(m^2+\Sigma_1)^{3/2}}=1
\end{equation}
Using the quantities
\begin{eqnarray}
m^2 &=&\frac{v^2}{4 \pi \overline{K} l_0^2} \mu^2
\label{eq:definition_mu} \\
\Sigma_1&
=&\frac{v^2}{4 \pi \overline{K} l_0^2} \sigma_1 \\
 \frac{4 \pi
\overline{K}}{v \delta} &=& \eta
\end{eqnarray}
Where $l_0$ and $d$ are defined respectively by
Eqs. (\ref{eq:definition_l0}) and (\ref{eq:definition_d}). The reduced
variable $\mu$ is defined in such way that The point at which the
replica symmetric solution becomes unstable has $\mu=1$.

the self-consistent equations are rewritten:
\begin{eqnarray}\label{eq:selfconsistent_rsb_adimensional}
\sigma_1&=&\frac 2 \eta \left[1-\exp(\eta\frac{\mu-1}{\mu})\right] \\
\mu^2&=&4 \left(\frac{l_0} d\right)^2 e^{
\frac{\eta(\mu-1)}{\mu}-\frac 1 {2\mu^3} e^{\eta\frac{\mu-1}{\mu}}}
\\ \mu^2+\sigma_1&=&1
\end{eqnarray}
  To solve
(\ref{eq:selfconsistent_rsb_adimensional}) we introduce $\varphi=\eta
\frac{\mu-1}{\mu}$. Physical solutions have $\eta>0$, $0\le \mu\le 1$ and
thus $\varphi\le 0$.  Excluding the solution $\mu=1$ from
(\ref{eq:selfconsistent_rsb_adimensional}) we obtain the following
equations in terms of $\varphi$:
\begin{eqnarray}
\mu(\varphi)&=&\sqrt{\frac 1 4 + 2\frac{e^\varphi -1}{\varphi}}-\frac
1 2 \label{eq:mu-of-phi} \\
4 \left(\frac{l_0} d\right)^2 &=& \mu^2(\varphi)
\exp\left[-\frac \varphi 2 + \frac{e^\varphi}{2
\mu^3(\varphi)}\right]=F(\varphi) \label{eq:F-of-phi}
\end{eqnarray}
We have thus reduced the self-consistent equations to a single equation
for $\varphi$. Putting $\varphi=0$ in
Eqs. (\ref{eq:mu-of-phi})--(\ref{eq:F-of-phi})  we
obtain $\mu=1$  and we recover the condition
(\ref{eq:condition_marginal}) on $d/l_0$, i.e. the limit of validity
of the RS solution. A plot of $F(\varphi)$ in shown in
Fig.~\ref{fig:f_of_phi}. As can be seen from this plot, $F$ has a
minimum for $\varphi=\varphi_c$. This implies that there can be no solution
of Eqs.  (\ref{eq:mu-of-phi})--(\ref{eq:F-of-phi}) when $l_0/d
<\sqrt{F(\varphi_c)}/2$. The physical values of $\varphi$ are thus located
in the interval $[\phi_c,0]$. Numerically, it is found that
$\varphi_c=-3.4325\pm 0.0025$ and $F(\varphi_c)=1.15338$. The corresponding
critical value of $l_0/d$ is then $l_0/d=0.536977$
i. e. $d/l_0=1.86...$.

\section{Functional Renormalization Group approach}\label{ap:frg_approach}

In this Appendix we detail the analysis using the Functional
Renormalization Group method of the effective model (\ref{eq:effec})
in presence of a mass term and correlated disorder in dimension $d$.
We use the notations of the classical equivalent model
(\ref{eq:acclassic}). The method is a Wilson momentum shell
integration which it is an extension of Ref.~\onlinecite{balents_loc}
to the case of a finite mass $m^2>0$.  Similar extensions can also be
found in
Ref.~\onlinecite{chauve_creep_short,chauve_creep_long,chauve_mbog_short,chauve_exact_rg}.

We start by studying $T_{cl}=0$. In the quantum problem this
corresponds to the limit $K\to 0$, $\hbar \to 0$, $\overline{K} =
K/\hbar $ fixed (see Eq.  (\ref{eq:corr_class_quant})). We consider
the ground state which is $\tau$ independent $\phi(x,\tau)=\phi(x)$.
It is in this limit that the GVM method revealed the presence of the
Mott Glass phase. It is more convenient to work with the function
$\Delta(\phi) = - R''(\phi)$, where the bare $R(\phi)$ has been
defined through (\ref{eq:quantumd}).  One first defines the running
dimensionless disorder:
\begin{eqnarray}
\tilde{\Delta}_l(\phi) = \frac{A_d}{c^2} \Lambda_l^{d-4}
\Delta_l(\phi)
\end{eqnarray}
with $A_d=S_d/(2 \pi)^d$, which is found to obey the $T_{cl}=0$ FRG
equation:
\begin{eqnarray} \label{eq:frg_flow_eq}
&& \partial_l \tilde{\Delta} = \epsilon \tilde{\Delta} - f(l) \left [
(\tilde{\Delta}^\prime)^2+ \tilde{\Delta}^{\prime
\prime}\left[\tilde{\Delta}-\tilde{\Delta}(0)\right]\right] \\ && f(l)
= \frac{1}{(1 + \mu e^{2 l})^{2}} \quad \mu=m^2/\Lambda^2
\end{eqnarray}
The UV cutoff is reduced to $\Lambda_l = \Lambda e^{-l}$ ($\Lambda
\sim 1/a$ where $a$ is lattice constant).  One can check that for
$m=0$, (\ref{eq:frg_flow_eq}) reduce to the one derived in
Ref.~\onlinecite{balents_loc}. This equation turns out to be identical
to the one describing point disorder in dimension $d$.

It is well known\cite{balents_loc} that in the case $m=0$ at $T=0$ a
cusp develops at the origin for $l\to \infty$. One finds that
$|\tilde{\Delta}(\phi,\infty)-\tilde{\Delta}(0,\infty)| \propto
|\phi|$ for $\phi \to 0$. This implies that $\lim_{l \to +\infty}
\tilde{\Delta}^{\prime \prime}(0,l)=-\infty$.  It is thus important as
a first step to analyze the cusp generation in the case $m \ne 0$. If
we define $\Delta_2(l)=- \tilde{\Delta}^{\prime \prime}(0,l)$, we have
from (\ref{eq:frg_flow_eq}):
\begin{equation}\label{eq:diff_eq_delta}
\partial_l \Delta_2= \epsilon \Delta_2 +f(l) \Delta_2^2
\end{equation}
This differential equation has for solution:
\begin{equation}
\frac 1 {\Delta_2(0)}- \frac{e^{\epsilon l}}{\Delta_2(l)}=\frac 1 2
\int_1^{e^{2l}} dx \frac 1 {x^{(d-2)/2}(1+\mu x)^2}
\end{equation}
Where $\Delta_2(0)$ is the bare disorder. Introducing the Larkin
length $R_c$ in the absence of a mass ($\mu=0$), defined as the length
scale at which $\Delta_2$ diverges i.e.
\begin{equation}
\frac 1 {\Delta_2(0)}=\frac 1 2 \int _1^{(R_c/a)^2}\frac
{dx}{x^{(d-2)/2}}
\end{equation}
One also obtains an equation that determines the Larkin length in the
presence of a mass $\tilde{R}_c(\mu)$ defined as the length where
$\Delta_2=+\infty$ for a non zero $\mu$ as a function of $R_c$ and
$\mu$ :
\begin{equation}\label{eq:larkin_length}
\int_1^{(\tilde{R}_c/a)^2}\frac{dx}{x^{(d-2)/2}(1+\mu x)^2}=\frac 2
{4-d}\left(\left(\frac {R_c} a\right)^{4-d}-1\right)
\end{equation}
(\ref{eq:larkin_length}) has two types of solutions, one with
$\tilde{R}_c=\infty$ for weak disorder and another one with
$\tilde{R}_c<\infty$ for stronger disorder . As discussed in the text,
this means that there are two phases, one in which disorder is strong
enough to generate a cusp and a second one in which the flow is cut by
the presence of the mass before a cusp can be generated. The former
corresponds to the Mott Glass phase, while the second one corresponds
to the Mott insulator phase. The equation of the transition line
between these two phases is obtained by setting $\tilde{R}_c=\infty$
in (\ref{eq:larkin_length}) and reads $R_c=R_c^*(\mu)$. At small $\mu$
and for $d<4$, we find that it behaves as:
\begin{equation}
\frac{R_c^*(\mu)}{a}\sim \frac{C(d)}{\sqrt{\mu}}
\end{equation}
where $C(d)$ is a dimension dependent constant.

The physical quantity which is directly affected by the presence of
the cusp is the tilt modulus $c_{44}(l)$. One finds that it satisfies
the RG flow equation:
\begin{eqnarray}
\partial_l \ln c_{44}(l) = - \Delta_2(l) f(l)
\label{c44}
\end{eqnarray}
while $c$ remains unrenormalized.  So clearly, either $\Delta_2(l)$
diverges sufficiently fast and $c_{44}(l=+\infty) = +\infty$
(interpreted as the Mott glass) or the mass cuts off the divergence
early enough and $c_{44}(l=+\infty)$ remains finite (Mott insulator).
>From the FRG it is possible to compute exactly the large $l$
behavior of the tilt modulus.  For that, and to make further progress
in the analysis of the two phases, we need to first consider the full
flow of the function $\tilde{\Delta}_l(\phi)$.  It can be shown easily
that the solution of the flow equations at $\mu \ne 0$ can be obtained
as a function of the solution at $\mu=0$ in the following way:
\begin{eqnarray}\label{eq:variable_change_T=0}
\tilde{\Delta}_{\mu}(\phi,l)=h(l)\tilde{\Delta}_{\mu=0}(\phi,t(l))\nonumber
\\ h(l)=\frac{e^{\epsilon l}}{1+\epsilon \int_0^l dl'\frac{e^{\epsilon
l'}}{(1+\mu e^{2l'})^2}} \nonumber \\ t(l)=\frac 1 \epsilon \ln \left(
1 +\epsilon \int_0^l \frac{e^{\epsilon l' dl'}}{(1+\mu
e^{2l'})^2}\right)
\end{eqnarray}
with the same initial condition. The behavior at large $l$ is the
following:
\begin{eqnarray}
&& h(l) \sim e^{\epsilon (l - l_c^*(\mu)) } \\ && t(l) \sim l_c^*(\mu)
\end{eqnarray}
where $R^*_c(\mu) = a e^{l_c^*(\mu)}$ was defined above. From
(\ref{c44}) it is then easy to see that $c_{44}(l) = c_{44}(0)
\Delta_2(l)/\Delta_2(0) e^{\epsilon l}$ which yields
e.g. $c_{44}(+\infty)$ in the no cusp phase as:
\begin{equation}
\frac{c_{44}(+\infty)}{c_{44}(0)}=\frac{\left(\frac {R_c}
a\right)^\epsilon -1}{\left( \frac {R_c} a \right)^\epsilon
-\left(\frac{R_c^*(\mu)}a \right)^\epsilon}
\end{equation}
One thus finds that the renormalized tilt modulus diverges as one
approaches the transition as:
\begin{equation}
c_{44}(+\infty) \sim (R_c-R_c^*(\mu))^{-1}
\end{equation}
In $d=4$ one has instead:
\begin{equation}
c_{44}(+\infty)=c_{44}(0) \frac{\ln(R_c/a)}{\ln(R_c/R_c^*(\mu))}
\end{equation}

In all cases one has $c_{44}(\infty) \to +\infty$ in the cusp
phase. On the contrary in the no cusp phase $c_{44}(\infty)$ remains
finite. We expect that having $c_{44}(\infty) \to +\infty$ leads to no
conductivity gap but having $c_{44}(\infty)<\infty$ produces a
conductivity gap.  In the cusp phase one can also expect that a term
$|\partial_z u|$ is generated \cite{balents_loc}. Such term give rise
to the transverse Meissner Effect. The critical field $h_{c1}$ needed
to bend vortices can be easily computed from the FRG.

We are now in position to estimate correlation functions in the case
of point disorder or their $z$ (i.e $\tau$) independent part in the
case of correlated disorder (i.e at $\omega_n=0$).  One has:
\begin{equation}
\overline{\langle\phi(q) \phi(-q)\rangle}=
\tilde{\Gamma}(q)=e^{dl}\Gamma(qe^l,\tilde{\Delta}(l),me^{2l})
\end{equation}

In the regime $q a \sim 1$, the correlation function $\Gamma$ can be
obtained by perturbation theory in $\tilde{\Delta}$.  Our strategy to
obtain correlation functions \cite{giamarchi_vortex_long} is therefore
to integrate the RG equations until $q a e^l \sim 1$. At this point,
we can calculate the correlation function $\Gamma$ perturbatively and
deduce $\tilde{\Gamma}$. We obtain:
\begin{equation}
\label{eq:correlation_T=0}
\tilde{\Gamma}(q)=\frac
{\tilde{\Delta}(0,l=\ln(1/aq))}{(aq)^{d-4}(q^2+m^2)^2}
\end{equation}

So that for $d=2$ one gets:
\begin{eqnarray}
\tilde \Gamma(q)=\frac C {q^2} \; \; q \gg m \nonumber \\ \tilde
\Gamma(q)=\frac {C'} {m^4} \; \; q \ll m
\end{eqnarray}
Note that the \emph{static} two point correlation functions or
equivalently the correlations for point disorder do not exhibit a
sharp transition.

It is crucial to check that the transition we found for $T_{cl}=0$
(i.e $\hbar=0$) survives at finite temperature (finite $\hbar$). In
the original quantum problem this corresponds to $K>0$ i.e. whether
the intermediate phase exists for interactions that are not infinitely
repulsive. The RG can be performed at finite $T_{cl}$.  Introducing
the effective running temperature:
\begin{eqnarray}
&& \tilde{T}_l = T_{cl} \frac{A_d \Lambda_l^{d-1}}{2 \sqrt{c c_z(l)} }
k(l) \\ && k(l) = (1 + \mu e^{2 l})^{-1/2}
\end{eqnarray}
one finds that the FRG equation becomes:
\begin{eqnarray}
&& \partial_l \tilde{\Delta} = \epsilon \tilde{\Delta} + \tilde{T}_l
\tilde{\Delta}'' - f(l) \left [ (\tilde{\Delta}^\prime)^2+
\tilde{\Delta}^{\prime
\prime}\left[\tilde{\Delta}-\tilde{\Delta}(0)\right]\right]
\end{eqnarray}
Note that in the quantum parameters $\tilde{T}_0 \sim K$.

In the absence of a mass, $\mu=0$, it is easy to see
\cite{chauve_mbog_short} that the temperature $\tilde{T}_l$ runs to
exactly zero at a finite length scale $l^*(\tilde{T}_0) - l_c \sim
c(d) \tilde{T}_0$ for small $\tilde{T}_0$ with $l_c =\ln (R_c/a)$ the
Larkin scale. This is because for $l>l_c$ the cusp is rounded
\cite{chauve_creep_short,chauve_creep_long} at finite $\tilde{T}_l$
with~:
\begin{eqnarray}
&& \Delta_2(l) \sim \frac{\Delta^{* \prime}(0^+)^2}{\tilde{T}_l} \sim
\chi \epsilon^2 \frac{1}{\tilde{T}_l}
\end{eqnarray}
where $\Delta^{*}(\phi)$ is the $T=0$ fixed point function, and $\chi$
a numerical constant.  Thus one can write:
\begin{eqnarray}
\partial_l \ln \tilde{T}_l &=& 1 - d - \frac{1}{2} \partial_l \ln
c_z(l) \\ & & = 1 - d - \chi \epsilon^2 \frac{1}{\tilde{T}_l}
\end{eqnarray}
This yield that $\partial_l \tilde{T}_l \approx - \chi \epsilon^2 $
and thus the temperature vanishes beyond the scale $l^*(\tilde{T}_0) -
l_c \sim \frac{1}{\chi \epsilon^2} \tilde{T}_0$.

It is thus clear that if $\mu$ is small enough so that
$l^*(\tilde{T}_0) \ll l_c^*(\mu)$ introduced above, the temperature
will vanish before the term $f(l)$ starts deviating from $1$ and
change the behavior of the solutions. Thus at small non zero
temperature the divergence of $c_{44}$ is not suppressed and the
transition survives.

A more detailed analytical study can be performed noticing that the
relation between the solution at finite $\mu$ and zero mass:
\begin{eqnarray}
\tilde{\Delta}_{\mu,\tilde{T}_l}(\phi,l)=h(l)\tilde{\Delta}_{\mu=0,\hat{T}_l}(\phi,t(l))
\end{eqnarray}
with the same functions $h(l)$ and $t(l)$ as above and $\hat{T}_l =
\tilde{T}_l/(h(l) f(l))$.  It confirms the above conclusions but will
not be detailed here.

Note finally that the above RG procedure uses that the thickness $L$
is constant. Since $\hbar$ runs to zero it means that $\beta$ runs to
infinity, and thus that the temperature is also irrelevant in the
quantum system.

%\bibliography{totphys,extra}

\end{document}